\documentclass{ws-jai}
\usepackage[flushleft]{threeparttable}
\usepackage{gensymb}
\makeatletter
\newcommand*{\rom}[1]{\expandafter\@slowromancap\romannumeral #1@}
\makeatother
\begin{document}

\catchline{}{}{}{}{} % Publisher's Area please ignore

\markboth{A. Gill, R. J. Selina, B. J. Butler}{A Study of the Compact Water Vapor Radiometer for Phase Calibration of the Karl G. Janksy Very Large Array}

\title{A Study of the Compact Water Vapor Radiometer for Phase Calibration of the Karl G. Janksy Very Large Array}

\author{Ajay Gill$^{1,2}$, Robert J. Selina$^{3}$, Bryan J. Butler$^{3}$}

\address{
$^{1}$National Radio Astronomy Observatory Graduate Summer Student, 1003 Lopezville Road, Socorro, NM, 87801, USA\\
$^{2}$Department of Astronomy and Astrophysics, University of Toronto, Toronto, ON, M5S 3H4, Canada\\
$^{3}$National Radio Astronomy Observatory, 1003 Lopezville Road, Socorro, NM, 87801, USA
}

\maketitle

\corres{$^{1}$Corresponding author: Ajay Gill (\texttt{ajay.gill@mail.utoronto.ca}) was a summer student at the National Radio Astronomy Observatory.}

\begin{history}
\received{(to be inserted by publisher)};
\revised{(to be inserted by publisher)};
\accepted{(to be inserted by publisher)};
\end{history}

\begin{abstract}
We report on laboratory test results of the Compact Water Vapor Radiometer (CWVR) prototype for the NSF's Karl G. Jansky Very Large Array (VLA), a five-channel design centered around the 22 GHz water vapor line. Fluctuations in precipitable water vapor cause fluctuations in atmospheric brightness emission, \textit{T$_{B}$}, which are assumed to be proportional to phase fluctuations of the astronomical signal, $\Delta\phi_{V}$, seen by an antenna. Water vapor radiometry consists of using a radiometer to measure variations in \textit{T$_{B}$} to correct for $\Delta\phi_{V}$. The CWVR channel isolation requirement of $<$ -20 dB is met, indicating $<$ 1\% power leakage between any two channels. Gain stability tests indicate that Channel 1 needs repair, and that the fluctuations in output counts for  Channel 2 to 5 are negatively correlated to the CWVR enclosure ambient temperature, with a change of $\sim$ 405 counts per 1$^{\circ}$ C change in temperature. With temperature correction, the single channel and channel difference gain stability is $<$ 2 $\times$ 10$^{-4}$, and the observable gain stability is $<$ 2.5 $\times$ 10$^{-4}$ over $\tau$ = 2.5 - 10$^3$ sec, all of which meet the requirements. Overall, the test results indicate that the CWVR meets specifications for dynamic range, channel isolation, and gain stability to be tested on an antenna. Future work consists of building more CWVRs and testing the phase correlations on the VLA antennas to evaluate the use of WVR for not only the VLA, but also the Next Generation Very Large Array (ngVLA).  
\end{abstract}


\section{Background}
\subsection{Phase fluctuations}
The prospect of water vapor radiometry (WVR) for correcting atmospheric phase fluctuations has long been recognized by the radio astronomy community. ALMA relies on radiometry of the 183 GHz water vapor line for observations at high frequencies \cite{Sterling2004}. The expanded bandwidth of the NSF's Karl G. Jansky Very Large Array (VLA) receivers and recent technological developments improve the prospects of WVR corrections for phase calibration for the VLA. Successful use of WVR for phase calibration at the VLA would also be a useful case study for using WVR corrections for the Next Generation Very Large Array (ngVLA).   

\begin{figure} [!htb]
\centering
\includegraphics[width=14cm, height = 6cm, keepaspectratio] {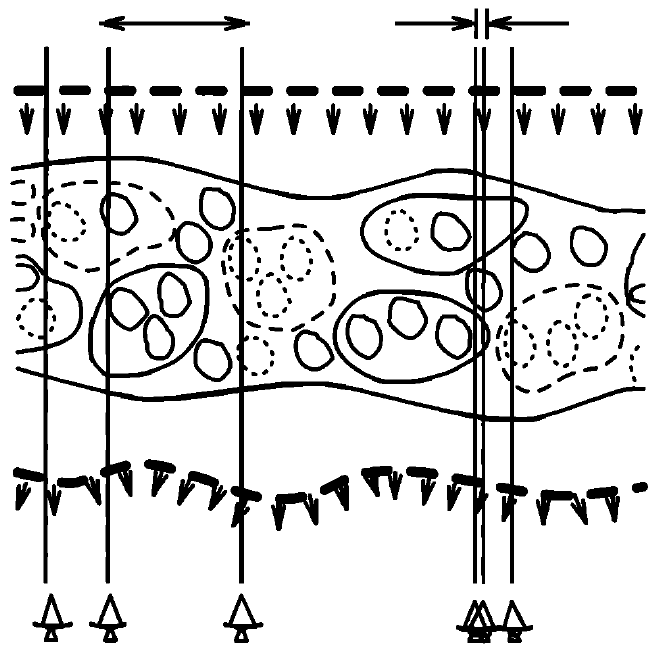}
\caption {Phase fluctuations due to variations in the index of refraction $n_{V}(\nu,t)$ (Desai, 1993).}
\label{PF}
\end{figure}

The troposphere is the lowest layer of the atmosphere, extending from the ground to an elevation of 7 to 10 km. The clouds and precipitable water vapor (PWV) move across the troposphere at $v_{V}$ $\sim$ 10 m$/$s, leading to a time and frequency varying refractive index, $n_{V}(\nu,t)$, in the layer. Planar radio wavefronts propagating through regions of $n_{V}(\nu,t)$ undergo refraction and an excess in electrical path length, leading to fluctuations of the phase of the wavefronts seen by an antenna, as shown in Figure \ref{PF}. 

The phase fluctuations due to PWV limit the spatial resolution of radio interferometers \cite{SH1997}. The total excess path length through the atmosphere can be estimated to be \cite{TS2001}
\begin{equation} \label{eqn1}
\mathcal{L} \simeq \mathcal{L}_{D} +  \mathcal{L}_{V} \simeq 0.228P_{0} + 6.3w \; \; 
\end{equation}

\noindent
where $\mathcal{L}_{D,V}$ is the excess path length due to dry air and PWV respectively, $P_{0}$  is the atmospheric pressure in millibars, and \emph{w} is the height of the column of water condensed from the atmosphere. We assume $P_{0}$ and $\mathcal{L}_{D}$ are slowly varying, and that the primary cause of phase fluctuations is $\mathcal{L}_{V}$. PWV of extent \textit{w} causes a phase change of the incoming radio wave, $\Delta\phi_{V}$, as \cite{B1999}

\begin{equation} \label{eqn2}
\Delta\phi_{V} \simeq \frac {12.6\pi} {\lambda} \times w  
\end{equation}

\begin{figure} [!htb]
\centering
\includegraphics[width=10cm, height = 7cm, keepaspectratio] {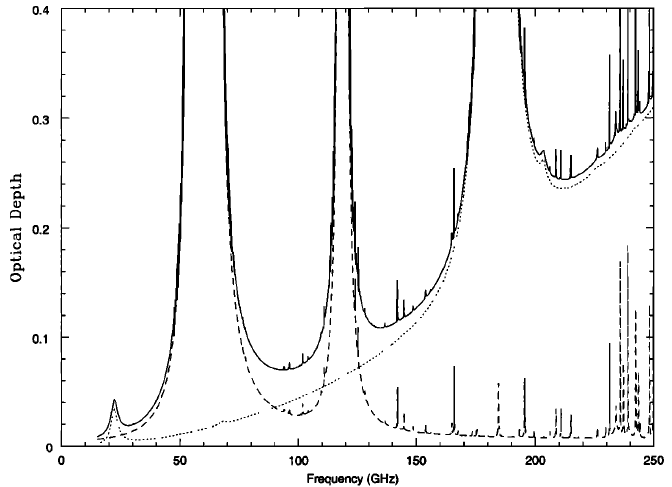}
\caption {Optical depth at the EVLA site with \textit{w} = 4 mm. The dotted line is the optical depth due to water vapor. The dashed line is the optical depth due to dry air ($O_{2}$ \& other trace gases). The solid line is the total optical depth \cite{Carilli}.}
\label{OD}
\end{figure}

A plot of the optical depth at the VLA site with \textit{w} = 4 mm is shown in Figure \ref{OD}. Below 130 GHz, both O$_{2}$ and H$_{2}$O contribute to the optical depth, whereas H$_{2}$O primarily contributes to the optical depth above 130 GHz. 

\subsection{Techniques for phase correction}
The various techniques used for phase calibration are briefly discussed below.
\begin{enumerate}
\item \emph{Self calibration} - The target source itself can be used for phase correction if it at least has a signal-to-noise ratio of $\sim$ 2 over the self-calibration averaging time \cite{CF1999}. This technique is not possible for weaker continuum or spectral line sources. 
\item \emph{Fast switching} - This technique consists of observing a strong calibrator source with a calibration cycle time short enough to reduce atmospheric phase fluctuations \cite{FP1999,HO1995}. Fast switching decreases observation time on the target source and places stringent constraints on the slew rate and mechanical settling time of the antenna. 
\item \emph{Paired array} - This technique uses a separate array that observes a nearby calibrator source while the target array continuously observes the target source \cite{Hold1992}. Paired elements constrain array configuration and reduce the sensitivity of the interferometer by a factor of 1.5 - 2 \cite{Clark2015}. 
\item \emph{Water vapor radiometry} - Fluctuations in PWV cause fluctuations in atmospheric brightness emission, \textit{T$_{B}$}, which are assumed to be proportional to phase fluctuations of the astronomical signal, $\Delta\phi_{V}$. WVR consists of using a radiometer to measure variations in \textit{T$_{B}$} to correct for $\Delta\phi_{V}$ \cite{Lay1997}. WVR must be able to work well consistently under a variety of weather conditions. 
\end{enumerate}

\subsection{Liquid water}

A complication arises for WVR because liquid and frozen water are also sources of continuum emission ($T_{B}$ $\propto$ $\nu$$^{2}$) but have a minimal contribution to the excess electrical path length. Therefore, water vapor emission may not correlate well with the astronomical phase fluctuations in the presence of liquid water, presenting a challenge for continuum radiometry. 

Water vapor emits both line and continuum emission. To distinguish water vapor emission from liquid water emission, one of the water vapor lines (22 or 183 GHz) can be observed with multiple channels around the line. One of the channels should be away from the water vapor emission line to distinguish the  $\nu$$^{2}$ continuum of liquid water from the vapor. WVR can be achieved in two ways: absolute and empirical radiometric phase correction.

\subsection{Absolute radiometric phase correction}

The brightness temperature due to atmospheric emission can be given by the radiometry equation as \cite{Dicke1946}

\begin{equation} \label{Dicke}
T_{B} = T_{atm} (1 - e^{-\tau_{tot}})
\end{equation}
\noindent where \emph{T}$_{atm}$ is the physical temperature of the atmosphere, and $\tau$$_{tot}$ is the total optical depth, which depends on dry air and water vapor. We assume $\tau$$_{tot}$ can be separated into three parts \cite{Carilli}

\begin{equation} \label{ttot}
\tau_{tot} = A_{\nu} w_{o} + B_{\nu} + A_{\nu} w_{rms}
\end{equation}
\noindent where \emph{A$_{\nu}$} is the optical depth per mm of PWV as a function of frequency, \emph{w$_{o}$} is the temporally stable component of PWV, \emph{B$_{\nu}$} is the temporally stable optical depth due to dry air as a function of frequency, and \emph{w$_{rms}$} is the time-varying component of PWV. We assume a temporally stable optical depth term, $\tau$$_{o}$ $\equiv$ \emph{A$_{\nu}$}\emph{w$_{o}$} + \emph{B$_{\nu}$}, and a time-varying optical depth term, $\tau$$_{rms}$ $\equiv$ \emph{A$_{\nu}$}\emph{w$_{rms}$}, and that $\tau$$_{o}$ $\gg$ $\tau$$_{rms}$. Inserting equation \ref{ttot} into \ref{Dicke} and assuming $\tau$$_{rms}$ $\ll$ 1 yields

\begin{equation} \label{Tbred}
T_{B} = T_{atm} (1 - e^{-\tau_{o}}) + T_{atm} e^{-\tau_{o}} \bigg[A_{\nu}w_{rms} - \frac {(A_{\nu}w_{rms})^2} {2} + ...\bigg]
\end{equation}

 The first term in equation \ref{Tbred} represents the temporally stable non-varying \emph{T$_{B}$}, and the second term represents the fluctuating component due to variations in PWV, which to first-order reduces to
\begin{equation} \label{Tbrms}
T^{rms}_{B} = T_{atm} e^{-\tau_{o}} A_{\nu}w_{rms}
\end{equation}

Absolute radiometric phase correction consists of measuring \textit{T}$^{rms}_{B}$ using a radiometer, inverting equation \ref{Tbrms} to derive the fluctuation in PWV, \emph{w$_{rms}$}, and using equation \ref{eqn2} to infer and correct for the fluctuation in phase, $\Delta\phi_{V}$, along the line of sight. However, there are some uncertainties involved with using absolute radiometry. The above derivations presume that the PWV fluctuations occur in a single narrow layer of known height, \emph{h$_{turb}$}, that remains constant over time. If the PWV fluctuations are instead distributed over a large range of altitudes, then the height of the dominant fluctuation at any given time must be known to convert \textit{T}$^{rms}_{B}$  into $\Delta\phi_{V}$. If PWV fluctuations at different heights contribute to \textit{T}$^{rms}_{B}$  simultaneously, the conversion becomes inaccurate. Another uncertainty in making absolute radiometric corrections involves errors in the theoretical atmospheric models that relate \emph{w} to \emph{T$_{B}$}.

\subsection{Empirical radiometric phase correction}

Absolute radiometric phase correction at each antenna places stringent requirements on the accuracy of the measured \emph{T$_{B}$}, ancillary data (\emph{T$_{atm}$}, \emph{h$_{turb}$}, etc.), and of the theoretical atmospheric models. A way to avoid some of these uncertainties is to empirically calculate the correlation between \textit{T}$^{rms}_{B}$ and $\Delta\phi_{V}$ by observing a strong calibration source at regular intervals. The phase fluctuation is a differential measurement between the phase, $\phi$, of antennas \emph{i} and \emph{j} of a single baseline taken as \cite{Chandlera}

\begin{equation} \label{diffP}
\Delta\phi_{V} = \phi_{i} - \phi_{j} 
\end{equation}

 The brightness emission fluctuation, \textit{T}$^{rms}_{B}$  $\equiv$ {\emph{$\Delta$T$_{B}$}}, to be compared with $\Delta$$\phi_{V}$ is also a differential measurement between the \emph{observable}, \emph{$\Delta$T}, of antennas \emph{i} and \emph{j} of the same baseline taken as

\begin{equation} \label{diffT}
\Delta T_{B} = \Delta T_{i} - \Delta T_{j} 
\end{equation}

 The scaling factor derived from the correlation between $\Delta$$T_{B}$ and $\Delta$$\phi_{V}$ can then be used to correct for $\Delta$$\phi_{V}$. Empirical radiometry consists of finding correlations between \emph{$\Delta$T$_{B}$} and $\Delta\phi_{V}$ for every baseline, so each antenna in the array requires a WVR. For preliminary tests with a single baseline, a minimum of two WVRs are required to test for correlations. A minimum of three WVRs are required to determine phase closure errors. 
 
 \section{Prior water vapor radiometers}
\begin{figure} [!htb]
\centering
\includegraphics[width=13cm, height=6cm, keepaspectratio]{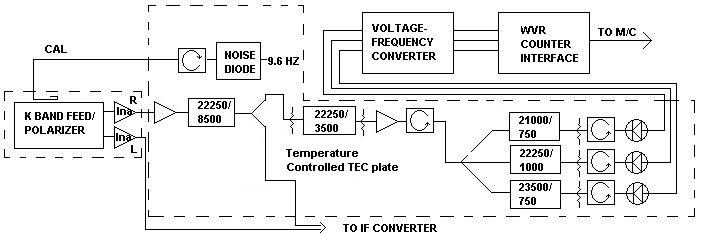}
\caption {VLA water vapor radiometer block diagram \cite{Chandlerb}.}
\label{WVRBD}
\end{figure}

A three-channel WVR was tested on VLA antennas 26 and 28 for one year as described in \cite{Chandlerb}. The 183 GHz line at the VLA site is too saturated, so the 22 GHz line was used. The block diagram of the VLA WVR is shown in Figure \ref{WVRBD}. 

\begin{figure} [!htb]
\centering
\includegraphics[width=3.5in]{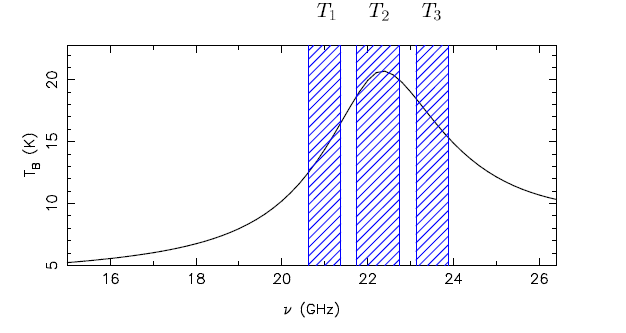}
\caption {Three channels of the VLA water vapor radiometer \cite{Chandlerb}.}
\label{3ch}
\end{figure}

The channels were placed at $\nu_{1}$ = 21.00 GHz with $\Delta\nu_{1}$ = 750.00 MHz, $\nu_{2}$ = 22.25 GHz with $\Delta\nu_{2}$ = 1000.00 MHz, and $\nu_{3}$ = 23.50 GHz with $\Delta\nu_{3}$ = 750.00 MHz.  The \emph{observable}, $\Delta T$, per antenna was defined as

\begin{equation} \label{eqn5}
\Delta T = w_{1}T_{1} + w_{2}T_{2} + w_{3}T_{3}
\end{equation}
\noindent
where the weights were \textit{$w_{1}$} = - 0.5, \textit{$w_{2}$} = 1.0, and \textit{$w_{3}$} = - 0.5, and \textit{$T_{1}$}, \textit{$T_{2}$}, and \textit{$T_{3}$} were the brightness temperatures of the three channels. The location of the channels is shown in Figure \ref{3ch}. 

For $\sim$ 35 $\mu$m of PWV, an excess path length $\mathcal{L}_{V}$ $\sim$ 220 $\mu$m or ($\lambda$/30 for $\lambda$ = 7 mm) was predicted from equation \ref{eqn1}. By considering possible atmospheric models and the channel locations, it was predicted that $\sim$ 35 $\mu$m of PWV would result in $\Delta$\textit{T} $\sim$ 25 mK. For an observable $\Delta T_{rms}$ $\sim$ 25 mK with the chosen weights, the stability of an individual channel needed to be $\sim$ 20 mK. For typical system temperatures of 50 - 100 K, this implied a gain stability $\Delta g/g$ $\sim$ 2 - 4 $\times$ 10$^-$$^4$. Test results showed that the sensitivity and gain stability requirements of the WVRs were met. The astronomical phase on baseline 26 - 28 was most significantly improved when the sky was clear and the phase fluctuations were large. Phase residuals were not significantly improved in the presence of clouds or when phase fluctuations were small. 

\section{Compact water vapor radiometer}

\subsection{Motivation}

The prior WVRs were designed for the VLA rather than the Expanded Very Large Array (EVLA). The K-band for the VLA had a frequency range from 21.40 - 24.40 GHz, so the VLA WVRs were limited to three channels because of the narrow 3.00 GHz bandwidth. The K-band for the EVLA has a frequency range of 18.00 - 26.50 GHz with a bandwidth of 8.50 GHz, so a five-channel compact water vapor radiometer (CWVR) was proposed, designed, and built. The five-channel CWVR improves the ability to distinguish liquid water from water vapor. The CWVR is based on Monolithic Microwave Integrated Circuit (MMIC) technology instead of discrete components, and its smaller size makes it easier to fit inside the space available in the vertex room of the antenna.

\subsection{CWVR system overview} 
\subsubsection{Block diagram}

\begin{figure} [!htb]
\centering
\includegraphics[width=6.5in]{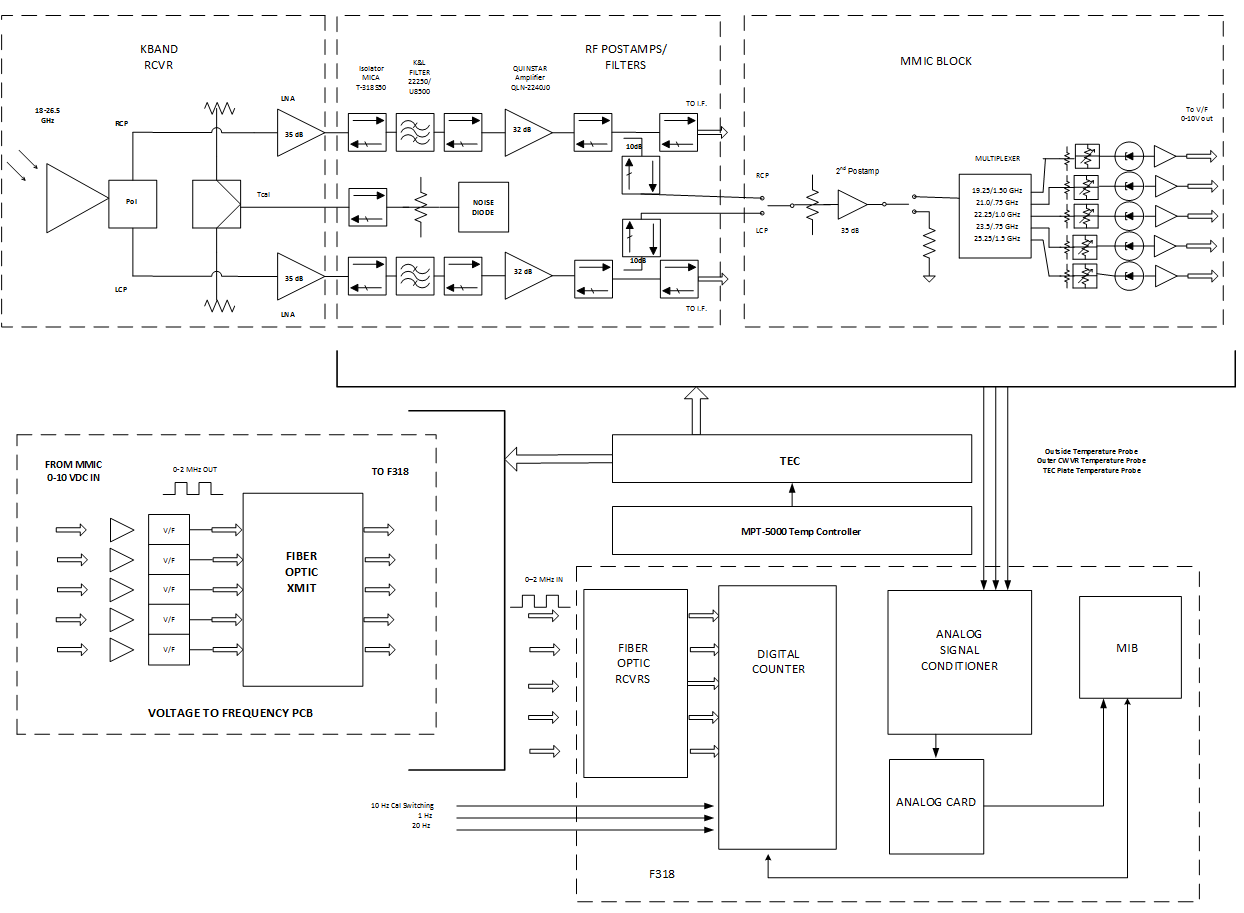}
\caption{Compact water vapor radiometer block diagram.}
\label{CWVRBD}
\end{figure}

The block diagram of the CWVR is shown in Figure \ref{CWVRBD}. Major elements of the diagram are described in the following sections. 

\subsubsection{K-band Dewar}

The astronomical signals arrive at the feed horn of the K-band receiver. The signals pass through a 90$\degree$ phase shifter, which converts the two linear orthogonal polarizations into two circular polarizations. The two circular polarizations are then split into right-hand circularly polarized (RCP) and left-hand circularly polarized (LCP) signals using an Orthomode Transducer. The calibration signal from the noise diode is coupled into the RCP and LCP signals. The RCP and LCP signals are then sent through low-noise amplifiers (LNAs) with a gain of 35 dB built by the NRAO Central Development Lab. The K-band Dewar is cooled to two stages of 15K and 50K to minimize noise generated by the LNA. Note that the noise diode is temperature stabilized as part of the CWVR retrofit.

\subsubsection{Isolators, post-amplifiers, and filters}

After an initial gain of 35 dB with the LNA, the RCP and LCP signals pass through a MICA T-318S50 isolator, a three-port device used to minimize reflections back into the input signal path by isolating the reflections to a third port. The signal from the isolator is sent to K\&L Microwave bandpass filters with a center frequency of 22.25 GHz and a bandwidth of 8.50 GHz, providing the 18.00 - 26.50 GHz bandwidth of the K-band receiver. After the filters, the signal passes through another MICA T-318S50 isolator, after which it goes through a Quinstar QLN-2240J0 post amplifier with a gain of 32 dB. After the post amp, the signal passes through another MICA T-318S50 isolator and a directional coupler. The through signal from the directional coupler labeled \emph{TO I.F} goes to the T303 IF down converter. The coupled signal is lower in power by 10 dB and goes to the MMIC block through another MICA T-318S50 isolator. This portion of the signal path is integrated into the CWVR as seen in Figure \ref{CWVRBD} and mounted on a temperature controlled plate for gain stability.

\subsubsection{MMIC block}

\begin{figure} [!htb]
\centering
\includegraphics[width=12cm, height = 7cm, keepaspectratio]{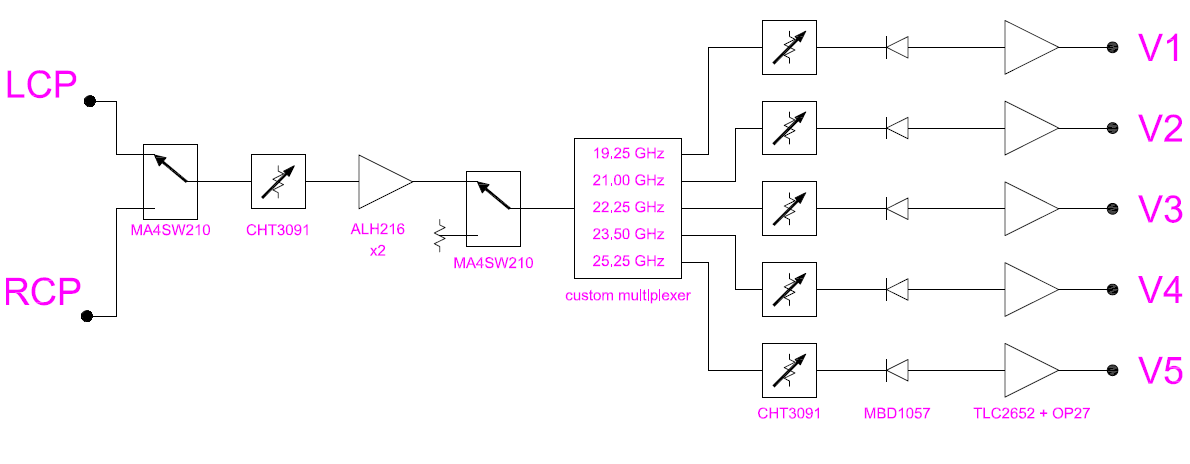}
\caption {Block diagram of the MMIC block.}
\label{MMIC}
\end{figure}

\begin{figure} [!htb]
\centering
\includegraphics[width=10cm, height = 6cm, keepaspectratio]{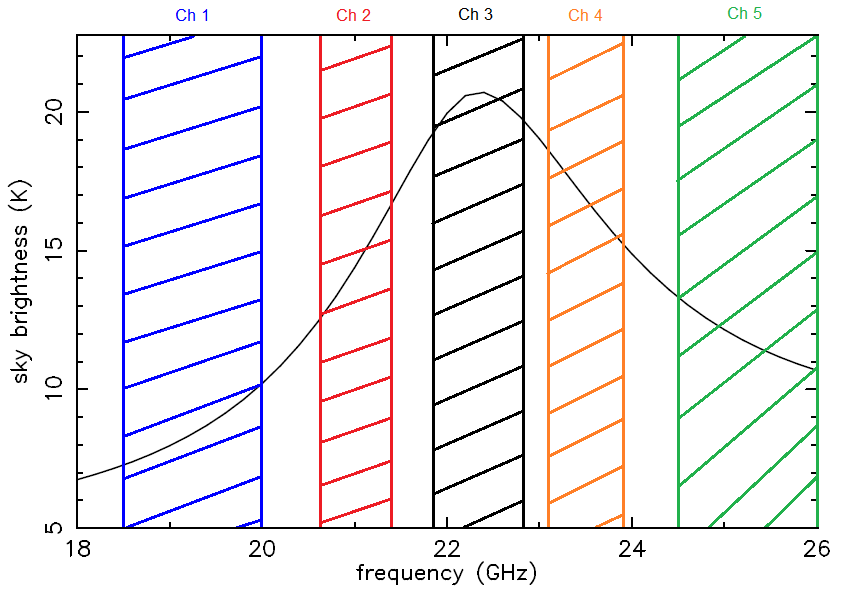}
\caption {Five channels of the CWVR.}
\label{5ch}
\end{figure}

The block diagram of the MMIC block is shown in Figure \ref{MMIC}. An input MA4SW210 PIN diode switch can select between the LCP and RCP input signals. The signal then passes through a CHT3091, which is a 100 step variable digital attenuator. The signal is amplified again by an ALH216 amplifier. The second MA4SW210 PIN diode switch can select between the incoming LCP/RCP signals and a termination to ground. The termination to ground allows DC offsets in the post-amps and diode detectors to be measured. The signal is then multiplexed with a frequency multiplexer into the 5 channels given in Table \ref{Channels} and Figure \ref{5ch}. 

\begin{wstable}[!htb]
\caption{Location of the 5 CWVR channels.}
\begin{tabular}{@{}ccccc@{}} \toprule
Channel & $\nu$$_{low}$ & $\nu$$_{center}$ &
$\nu$$_{high}$ & $\Delta\nu$ \\
& (GHz) & (GHz) & (GHz) & (GHz) \\ \colrule
1                & 18.500                     & 19.250                        & 20.000                      & 1.500                      \\ 
2                & 20.625                     & 21.000                        & 21.375                      & 0.750                      \\ 
3                & 21.750                     & 22.250                        & 22.750                      & 1.000                      \\ 
4                & 23.125                     & 23.500                        & 23.875                      & 0.750                      \\ 
5                & 24.500                     & 25.250                        & 26.000                      & 1.500                      \\ \botrule
\end{tabular}
\label{Channels}
\end{wstable}

The DC offset counts were measured for a period of 90 minutes, and the median values per channel are given in Table \ref{Offset}.

\begin{wstable}[!htb]
\caption{Median DC offset counts per channel.}
\begin{tabular}{@{}cccccc@{}} \toprule
 & Ch 1 & Ch 2 &
Ch 3 & Ch 4 & Ch 5 \\ \colrule
Mean offset counts & 272 & 270 & 296 & 235 & 207 \\ \botrule
\end{tabular}
\label{Offset}
\end{wstable}

The signal then goes through another CHT3091 attenuator per channel. The attenuators in the MMIC block are controlled through the function control box with a DB25 cable. Then, the signal goes to a MBD1057 tunnel diode detector, with the detector voltage output amplified by TLC2652 and OP27 amplifiers. The 0 - 10 V output from the amplifiers, proportional to the input RF power per channel, is then sent to the Voltage to Frequency board.

\subsubsection{Voltage to Frequency board}

The 0 - 10V output per channel from the MMIC block is sent to a Voltage to Frequency (V-F) Converter (VFC110AP), which converts the 0 - 10 V voltage inputs to 0 - 2 MHz frequency outputs. The frequency outputs of the five channels are sent to the F318 module using ST-E2000 fiber optic cables. 

\subsubsection{F318 module}  
\begin{figure} [!htb]
\centering
\includegraphics[width=9cm, height = 8cm, keepaspectratio]{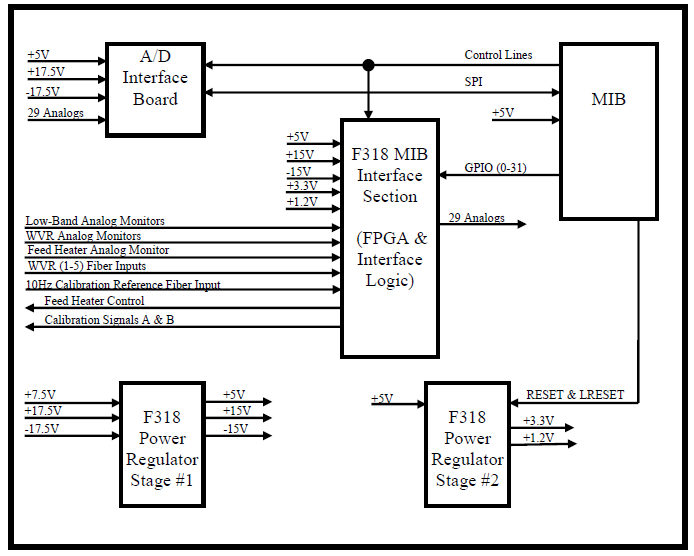}
\caption {F318 module block diagram (Koski, 2017).}
\label{F318BD}
\end{figure}

 The block diagram of the F318 module is shown in Figure \ref{F318BD}. The F318 module is required to interface the EVLA monitor and control system to the fiber optic cables of the five channel CWVR. The F318 also monitors temperatures within the CWVR. Figure \ref{F318BD} shows the main boards of the module. The Analog to Digital Interface Board brings in a total of twenty-nine analog voltages through four input channels. The F318 MIB Interface Board is the primary interface between the MIB and the CWVR, with the frequency counter implemented in the Xilinx FPGA. These two boards along with the MIB provide for a simple interface for analog and digital monitoring of the CWVR. Further details on the functionality of the F318 module with the CWVR are provided in \cite{Koski2017}.
 
\begin{figure} [!htb]
\centering
\includegraphics[width=13cm, height = 7cm, keepaspectratio]{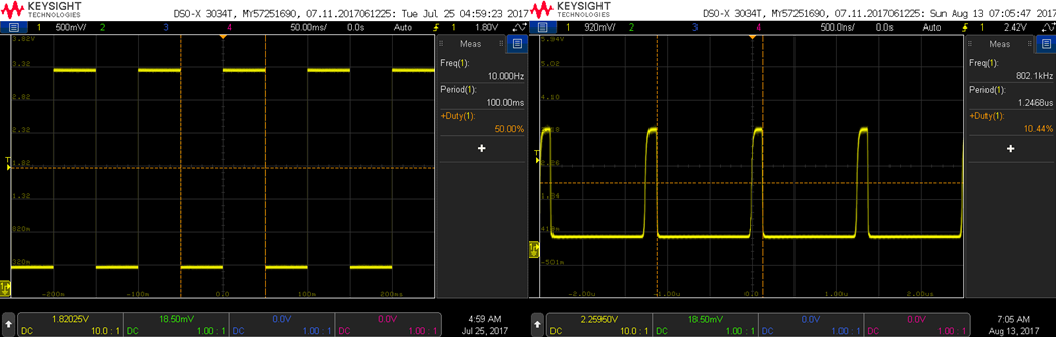}
\caption {10 Hz calibration signal (left) $\&$ channel frequency output (right).}
\label{VFWave}
\end{figure}

The 10 Hz noise diode calibration signal, with a period of 100 ms and 50$\%$ duty cycle, is used to count the number of pulses every high and low cal cycle. The data is collected every 100 ms at the rising edge of the 10 Hz signal. For the waveforms shown in Figure \ref{VFWave}, as an example

\begin{equation}
Ch\emph{1}_{low,high}= \frac {802.1 \times 10^3 \; counts} {sec} \times 50 \; ms \times \frac {1 \; sec} {10^3 \; ms} \simeq 40,000 \; counts
\end{equation}
\begin{equation}
Ch\emph{1}_{total}= Ch\emph{1}_{low} + Ch\emph{1}_{high} \simeq 80,000 \;counts
\end{equation}

\subsubsection{Temperature control}

The temperature of the temperature controlled plate (TCP) on which the post-amps, filters, and the MMIC block are mounted is maintained using a CP-036 Peltier Thermoelectric Cooler (TEC). The TEC is controlled using a MPT-5000 linear, bipolar temperature controller. The MPT-5000 contains a 12 turn trimpot, which can be adjusted to set the desired temperature. The heat from the TCP is absorbed into the heat sink of the TEC, and a fan is used to remove this heat to the outside of the CWVR. For laboratory testing, the temperature of the TCP was set to 30\degree C to minimize reflections and losses in Teflon cables. The PID control loop employs a TCS-620 temperature sensor, which has a resistance of 20 k$\ohm$ at 25\degree C. 

The Temperature Setpoint (TS) and Temperature Monitor (TM) signals of the MPT-5000 are used to set the temperature of the TEC and to monitor the temperature of the TCS-620, respectively. During stable operation, the voltage of TS closely matches that of TM. Two additional AD590 temperature sensors were used to monitor the temperature of the TCP adjacent to the post amps and the ambient temperature within the CWVR enclosure. Table \ref{MIBlabels} lists the signal names and descriptions of signals on the MIB for signals related to temperature control and monitor.  

\begin{wstable}[!htb]
\caption{Signal names on Monitor and Interface Board.}
\begin{tabular}{@{}cc@{}} \toprule
Signal name & Description\\ \colrule
WVR 7 & Temperature monitor for TCS-620 (TCP)\\
WVR 8 & Temperature setpoint for TEC \\
WVR\_Temp1 & Temperature monitor for AD590 (CWVR ambient)\\
WVR\_Temp2 & Temperature monitor for AD590 (TCP)\\
\botrule
\end{tabular}
\label{MIBlabels}
\end{wstable}

\section{Compact water vapor radiometer requirements}

\subsection{Channel isolation requirement}

The integrated isolation between any two channels is defined as

\begin{equation} \label{Iso}
IS_{xy}= \frac {\int_{\nu_{i}}^{\nu_{f}} P_{x}(\nu) P_{y}(\nu) d\nu} {\int_{\nu_{i}}^{\nu_{f}} P_{y}(\nu) d\nu}
\end{equation}
\noindent where $IS_{xy}$ is the power leakage from channel x into channel y, $\nu_{i}$ = 17.00 GHz, $\nu_{f}$ = 27.00 GHz, and $P_{x}$($\nu$) and $P_{y}$($\nu$) is the output power at frequency $\nu$ of channel x and y, respectively. The requirement for channel isolation $IS_{xy}$ was specified to be 	$<$ -20 dB, or a power leakage from channel x into channel y of $<$ 1$\%$.

\subsection{Gain stability requirement}
From the initial stability tests, Channel 1 was measured to be defective, so it was subsequently ignored, and the gain stability requirements were set for Channels 2 to 5. For Channels 2 to 5 of the CWVR given in Table \ref{Channels}, the \emph{observable}, $\Delta T$, per antenna can be defined as

\begin{equation} \label{obs}
\Delta T \propto \Delta P_{in} = w_{2}P_{2} + w_{3}P_{3} + w_{4}P_{4} + w_{5}P_{5}
\end{equation}
\noindent
where the weights are typically \textit{$w_{2}$} = -0.5, \textit{$w_{3}$} = 1.0, \textit{$w_{4}$} = -0.5, \textit{$w_{5}$} = 0.25, and \textit{$P_{2}$}, \textit{$P_{3}$}, \textit{$P_{4}$}, and \textit{$P_{5}$} are the power outputs due to sky emission in the five channels. It can be estimated from equation \ref{eqn1} that $\sim$ 35 $\mu$m of PWV leads to $\mathcal{L}_{V}$ $\sim$ 220 $\mu$m of excess electrical path length ($\lambda$/30 for $\lambda$ = 7 mm) and will result in $\Delta$$T$ $\sim$ 25 mK \cite{B1999}. For $\Delta T_{rms}$ $\sim$ 25 mK, the stability of each individual channel needs to be $\sim$ 20 mK. For typical system temperatures of 50 - 100 K, this requires a gain stability for $\Delta g/g$ $\sim$ 2.5 - 5 $\times$ 10$^-$$^4$ for $\Delta T$ and $\Delta$$g_{i}$$/$$g_{i}$ $\sim$ 2 - 4 $\times$ 10$^-$$^4$ for each individual channel. 

 The short timescale, $\tau_{short}$, for which the CWVR needs to be gain stable is

\begin{equation} \label{taushort}
\tau_{short} = \frac {D} {v_{V}} = \frac {25 \; m} {10 \; m/sec} = 2.5 \; sec
\end{equation}
\noindent where \textit{D} is the diameter of the antenna and $v_{V}$ is the average speed of water vapor across the troposphere. The long timescale for which the CWVR needs to be gain stable is $\tau_{long}$ $\sim$ 10$^3$ sec to allow for longer calibration cycles of the interferometer. The stability of the CWVR channels on different timescales, $\tau$, was characterized by the Allan Standard Deviation (ASD), $\sigma$$_{P}$($\tau$) defined by

\begin{equation} \label{ASD}
\sigma_{P}(\tau) = \left\{\frac {1} {2} \langle \left\{P(t) - P(t-\tau)\right\}^2 \rangle    \right\}^\frac {1} {2}
\end{equation}

 For single channel ASDs with N time and output power data points, equation \ref{ASD} becomes

\begin{equation} \label{ASDSC}
\sigma_{P}(\tau) = \left\{\frac {1} {2} \sum_{j=1}^{N/2} \frac {1} {N - \tau_{j}} \sum_{i=1}^{N - \tau_{j}}  \left\{P(t_{i}+\tau_{j}) - P(t_{i})\right\}^2   \right\}^\frac {1} {2}
\end{equation}
\noindent where $\sigma_{P}$($\tau$) is normalized with respect to

\begin{equation} \label{MSC}
\mu_{P} = \langle P(t) \rangle
\end{equation}

 For channel difference ASDs with N time and output power data points, equation \ref{ASD} becomes

\begin{equation} \label{ASDDC}
\sigma_{P_{x}-P_{y}}(\tau) = \left\{\frac {1} {2} \sum_{j=1}^{N/2} \frac {1} {N - \tau_{j}} \sum_{i=1}^{N - \tau_{j}}  \left\{[P_{x}(t_{i}+\tau_{j}) - P_{y}(t_{i}+\tau_{j})] - [P_{x}(t_{i}) - P_{y}(t_{i})]       \right\}^2   \right\}^\frac {1} {2}
\end{equation}
\noindent where $\sigma_{P_{x}-P_{y}}$($\tau$) is normalized with respect to

\begin{equation} \label{MDC}
\mu_{P_{x}-P_{y}} = \frac {\langle P_{x}(t) + P_{y}(t) \rangle} {2} 
\end{equation}

 For the ASD of $\Delta$$P_{in}$ with N time and output power data points, equation \ref{ASD} becomes

\begin{equation} \label{ASDobs}
\sigma_{\Delta P_{in}}(\tau) = \left\{\frac {1} {2} \sum_{j=1}^{N/2} \frac {1} {N - \tau_{j}} \sum_{i=1}^{N - \tau_{j}}  \left\{\sum_{k=2}^{5} w_{k}P_{k}(t_{i} + \tau_{j}) - \sum_{k=2}^{5} w_{k}P_{k}(t_{i})        \right\}^2   \right\}^\frac {1} {2}
\end{equation}
\noindent where $\sigma_{\Delta P_{in}}$($\tau$) is normalized with respect to

\begin{equation} \label{Mobs}
\mu_{\Delta P_{in}} = \frac {\langle P_{2}(t) + P_{3}(t) + P_{4}(t) + P_{5}(t)\rangle} {4} 
\end{equation}

\subsection{Temperature stability requirement}
Since the temperature coefficients of the CWVR components were not established prior to these tests, the physical temperature stability requirement inside the CWVR and on the TCP was arbitrarily set to $\sim$ 25 mK over 10$^3$ sec timescales. 

\section{Results}
\subsection{Dynamic range}
\begin{figure} [!htb]
\centering
\includegraphics[width=20cm, height = 10cm, keepaspectratio]{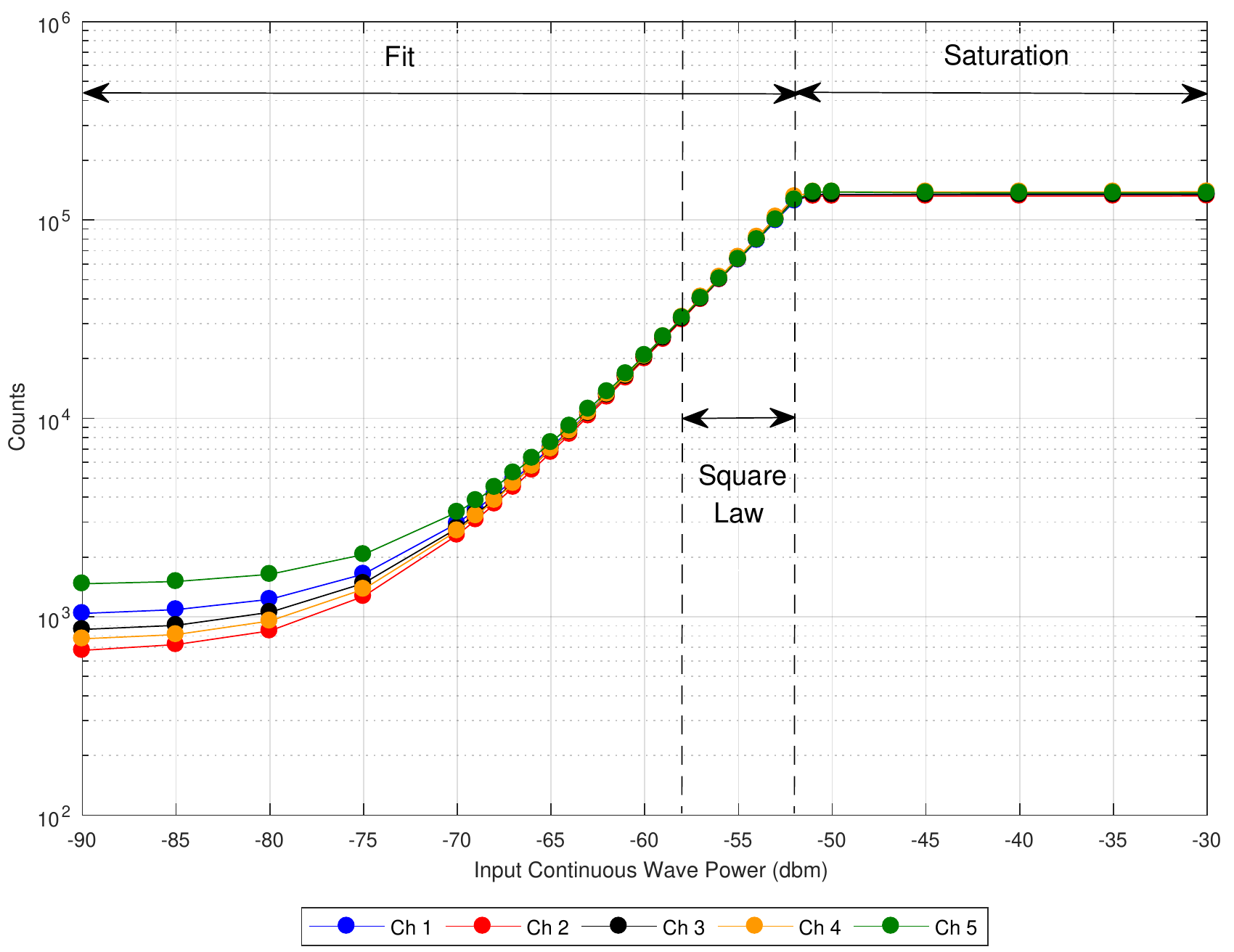}
\caption {Dynamic range of the diode detectors.}
\label{DRange}
\end{figure}

The dynamic range of the MBD1057 tunnel diode detectors was determined by sending input continuous wave power at the center of each channel at power levels ranging from -90 to -30 dbm. The resulting plot of output counts versus input power level per channel in shown in Figure \ref{DRange}. At -90 dbm input, the channels are at the noise floor. Channel 5 has the highest noise floor possibly due to its wider than desired design bandwidth.

As the input power increases, the counts increase and approach the linear region labeled \emph{square law}. In the square-law region, \emph{$\Delta$T} $\propto$ $\Delta$$P_{in}$ $\propto$ $V_{out}$ $\propto$ $Counts_{out}$. The CWVR should be operated in the square-law region, as it is in this region that the output counts are most sensitive to changes in input power. The square-law region ranges from $\sim$ -58 to -52 dbm, providing a dynamic range of $\sim$ 6 dbm. At input powers above -52 dbm, the diode detectors go into saturation.  

In the region labeled \emph{fit} ranging from -90 to -52 dbm, the curve for each channel was fit to the form given in equation \ref{Fit} to find the relationship between counts and input power. The fit effectively increases the dynamic range of the detectors to $\sim$ 18 dbm. 

\begin{equation} \label{Fit}
P_{in} (dbm) =\frac {1} {A} \left\{\ln(Counts) - \ln(B)\right\}
\end{equation}
\noindent where the constants A and B per channel are given  in Table \ref{Constants}. 

\begin{wstable}[!htb]
\caption{Constants per channel for equation \ref{Fit}.}
\begin{tabular}{@{}ccc@{}} \toprule
& A & B \\ \colrule
Ch 1 & 0.225 & 1.518$\times$$10^{10}$ \\ 
Ch 2 & 0.232 & 2.191$\times$$10^{10}$ \\
Ch 3 & 0.228 & 1.809$\times$$10^{10}$ \\ 
Ch 4 & 0.231 & 2.107$\times$$10^{10}$ \\ 
Ch 5 & 0.224 & 1.455$\times$$10^{10}$ \\
\botrule
\end{tabular}
\label{Constants}
\end{wstable}

\subsection{Channel isolation}
\begin{figure} [!htb]
\centering
\includegraphics[width=15cm, height = 10cm, keepaspectratio]{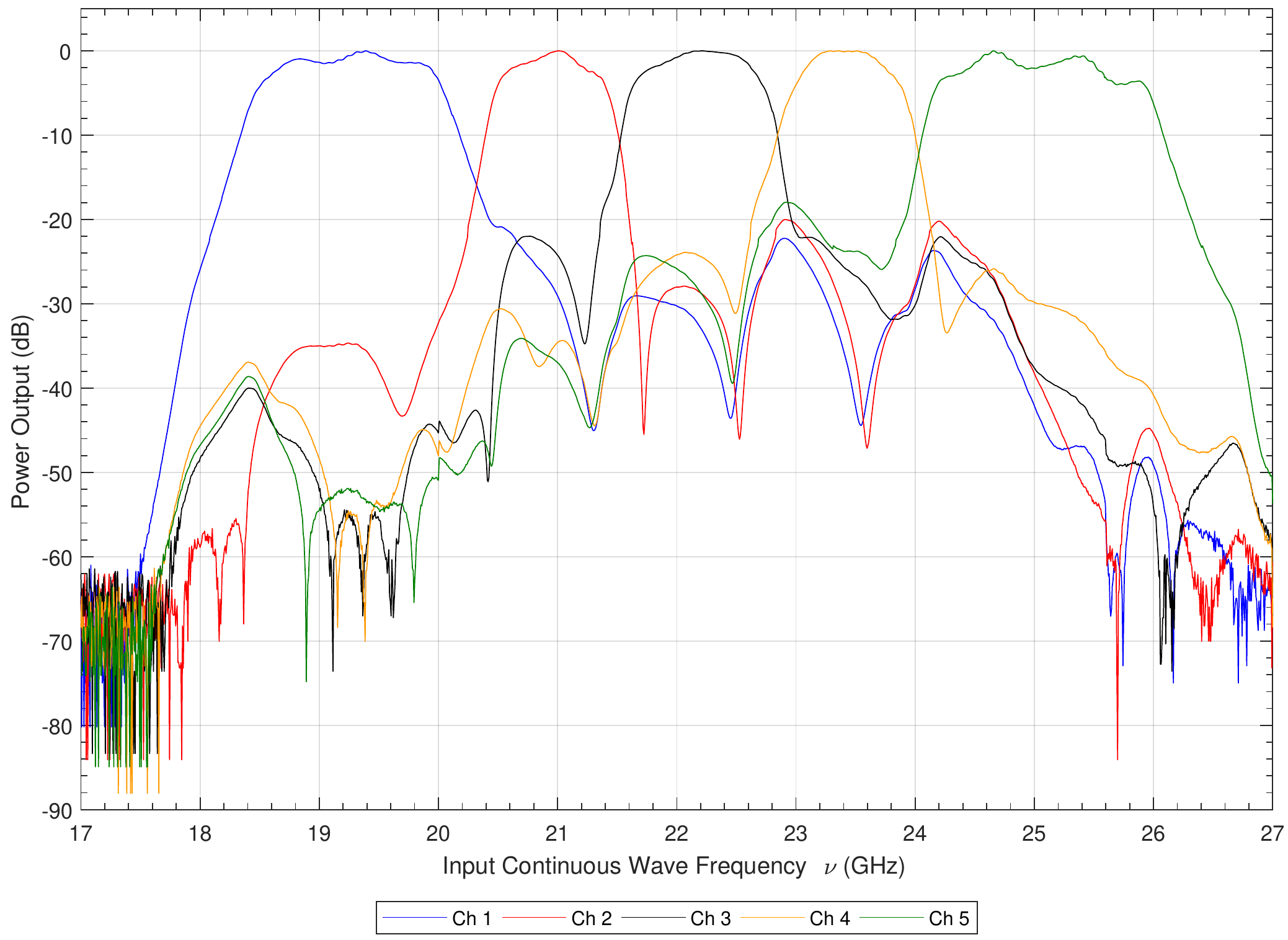}
\caption {Power response of the CWVR channels as a function of frequency.}
\label{Presponse}
\end{figure}

The power response of the CWVR was measured as a function of frequency. Frequency sweeps were done from 17.00 to 27.00 GHz at $\sim$ 6.41 MHz/sec at input power levels from -30 to -55 dbm with increments of -5 dbm. The Keysight PSG E8257D 250.00 kHz - 40.00 GHz Analog Signal Generator was used for the sweeps, and a MegaPhase RF Orange cable with a maximum operating frequency of 50.00 GHz was used to connect the generator to the CWVR input. For the sweeps, input power was sent only to the LCP input of the CWVR. The overall power response was determined by combining the unsaturated regions of the response curves at input power levels ranging from -30 to -55 dbm, and it is shown as a function of frequency in Figure \ref{Presponse}.

Using the data in Figure \ref{Presponse}, the isolation between all channels was calculated using equation \ref{Iso} and the results are given in Table \ref{IsoTable}. All  values are $<$ -20 dB, indicating $<$ 1$\%$ power leakage between any two channels, which meets the specification. 

\begin{wstable}[!htb]
\caption{Measured isolation between CWVR channels.}
\begin{tabular}{@{}cccc@{}} \toprule
\boldmath{$IS_{12}  (dB)$} & \boldmath{$IS_{13}  (dB)$} & \boldmath{$IS_{14}  (dB)$} & \boldmath{$IS_{15}  (dB)$}  \\ \hline
 -25.09               & -30.79               & -29.25               & -32.38               \\ \hline
\boldmath{$IS_{21}  (dB)$} & \boldmath{$IS_{23}  (dB)$} & \boldmath{$IS_{24}  (dB)$} & \boldmath{$IS_{25}  (dB)$}  \\ \hline
-27.42               & -22.90               & -26.10               & -27.72               \\ \hline
\boldmath{$IS_{31}  (dB)$} & \boldmath{$IS_{32}  (dB)$} & \boldmath{$IS_{34}  (dB)$} & \boldmath{$IS_{35}  (dB)$}  \\ \hline
-31.99               & -21.78               & -20.25               & -25.46               \\ \hline
\boldmath{$IS_{41}  (dB)$} & \boldmath{$IS_{42}  (dB)$} & \boldmath{$IS_{43}  (dB)$} & \boldmath{$IS_{44}  (dB)$}  \\ \hline
-31.15               & -25.67               & -20.95               & -22.74               \\ \hline
\boldmath{$IS_{51}  (dB)$} & \boldmath{$IS_{52}  (dB)$} & \boldmath{$IS_{53}  (dB)$} & \boldmath{$IS_{54}  (dB)$}  \\ \hline
-32.34               & -25.35               & -24.22               & -20.81               \\ \hline
\end{tabular}
\label{IsoTable}
\end{wstable}

\subsection{Gain stability}
\begin{figure} [!htb]
\centering
\includegraphics[width=12cm, height = 7cm, keepaspectratio]{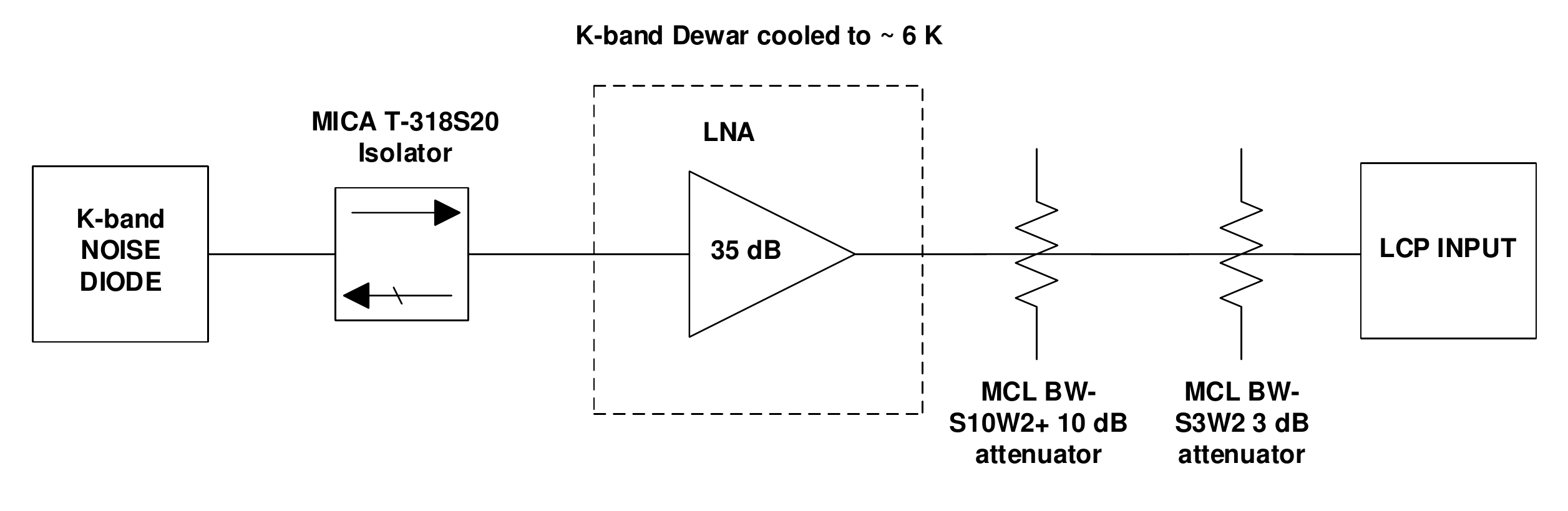}
\caption {Block diagram of gain stability test setup.}
\label{GSBlock}
\end{figure}

The instrumental setup used to test the gain stability of the CWVR is shown in Figure \ref{GSBlock}. The K-band noise diode source from within the CWVR generates broadband noise from 18.00 - 26.50 GHz. The noise diode signal is taken as an output from the CWVR and is sent through a MICA T-318S20 isolator to minimize reflections. The noise diode signal with no amplification to weak to provide output counts in the square-law region, so a K-band LNA with a gain of 35 dB was used to amplify the signal. The LNA is within the Dewar, which is cooled to $\sim$ 6 K to minimize noise and gain fluctuations generated by the LNA itself. The output signal from the Dewar has 13 dB of attenuation before going to the LCP input of the CWVR. 

With the combination of the 13 dB attenuation and adjustment of the CHT3091 attenuators in the MMIC block, the output counts were set to $\sim$ 60,000 counts per channel in the square-law region of the diode detectors. The test was run for a period of 64 hours, and data of the output counts per channel and the temperature sensors was collected every second. 

\begin{figure} [!htb]
\centering
\includegraphics[width=12cm, height = 5cm, keepaspectratio]{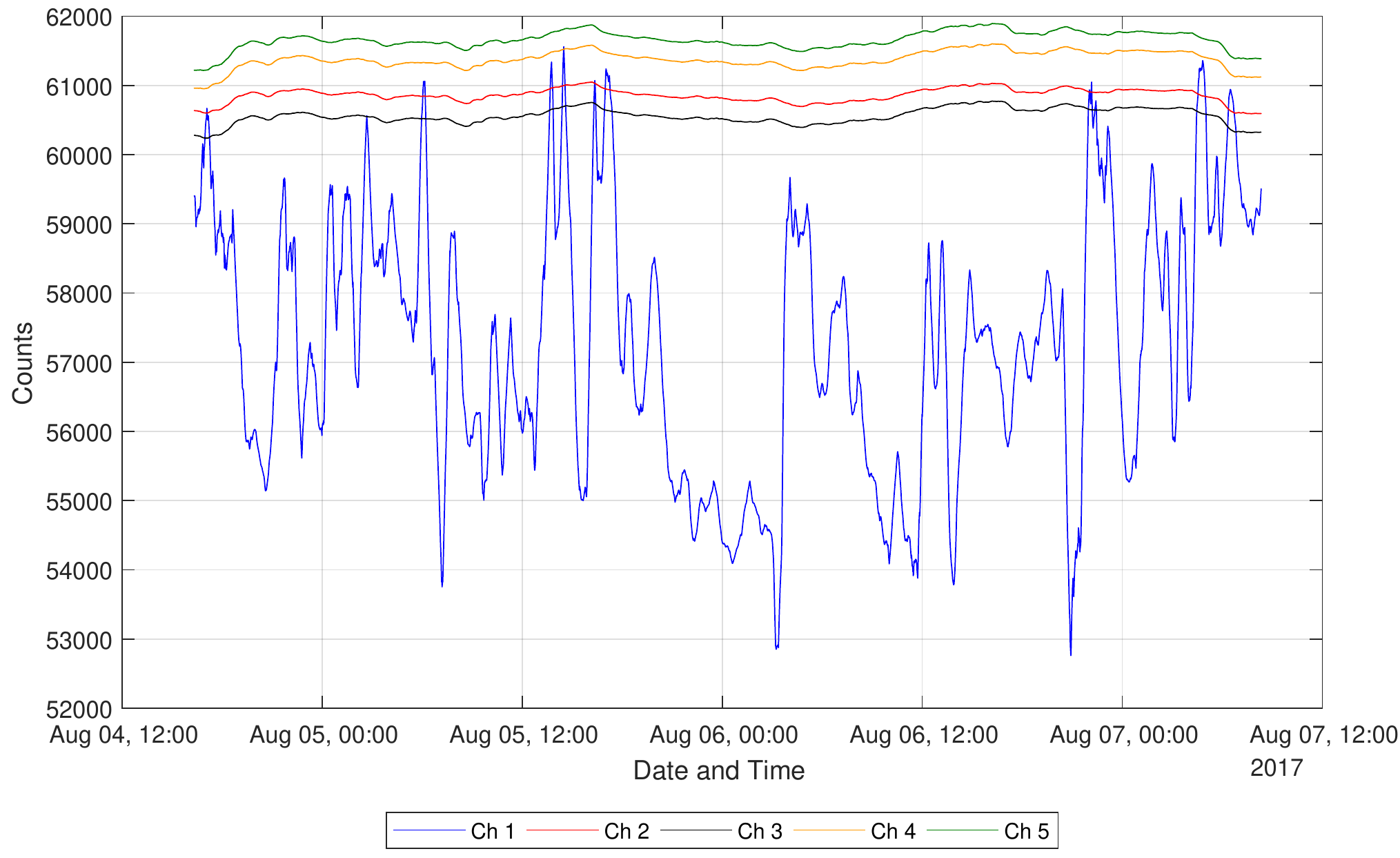}
\caption {Output counts of Channel 1 to 5 over 64 hrs averaged with 20 min running mean.}
\label{Ch1-5}
\end{figure}

\begin{figure} [!htb]
\centering
\includegraphics[width=12cm, height = 5cm, keepaspectratio]{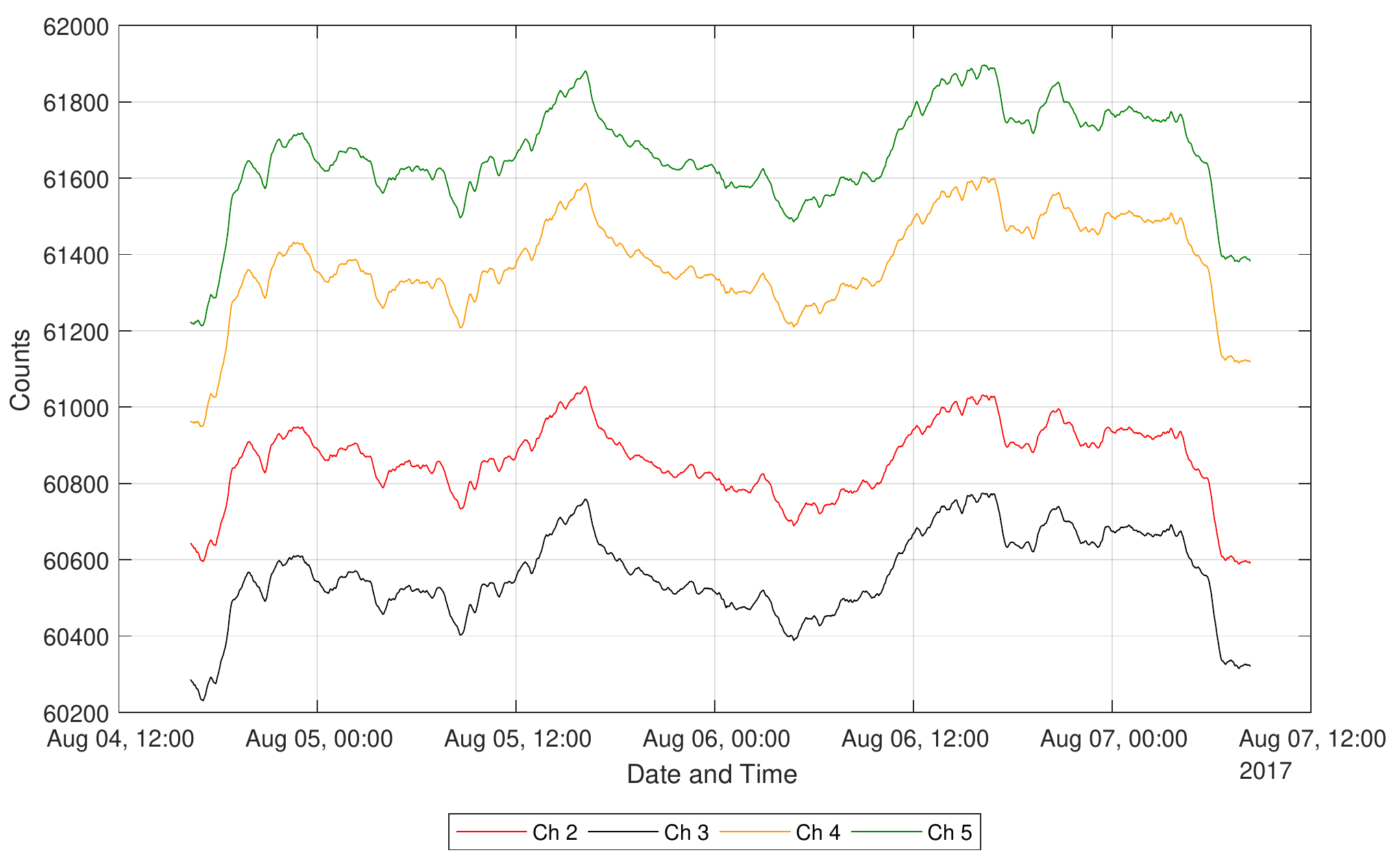}
\caption {Output counts of Channel 2 to 5 over 64 hrs averaged with 20 min running mean.}
\label{Ch2-5}
\end{figure}

The measured output counts per channel over the 64 hr period averaged with a 20 min running mean are shown in Figure \ref{Ch1-5}. Channel 1 is much more unstable than the rest, possibly due to its attenuation varying with time, so it was subsequently ignored. The counts from Channel 2 to 5 averaged with a 20 min running mean are shown in Figure \ref{Ch2-5}. There is a slight offset in counts between the channels due to the step size of the CHT3091 attenuators.

\subsubsection{Single channel Allan Standard Deviations}
\begin{figure} [!htb]
\centering
\includegraphics[width=12cm, height = 6cm, keepaspectratio]{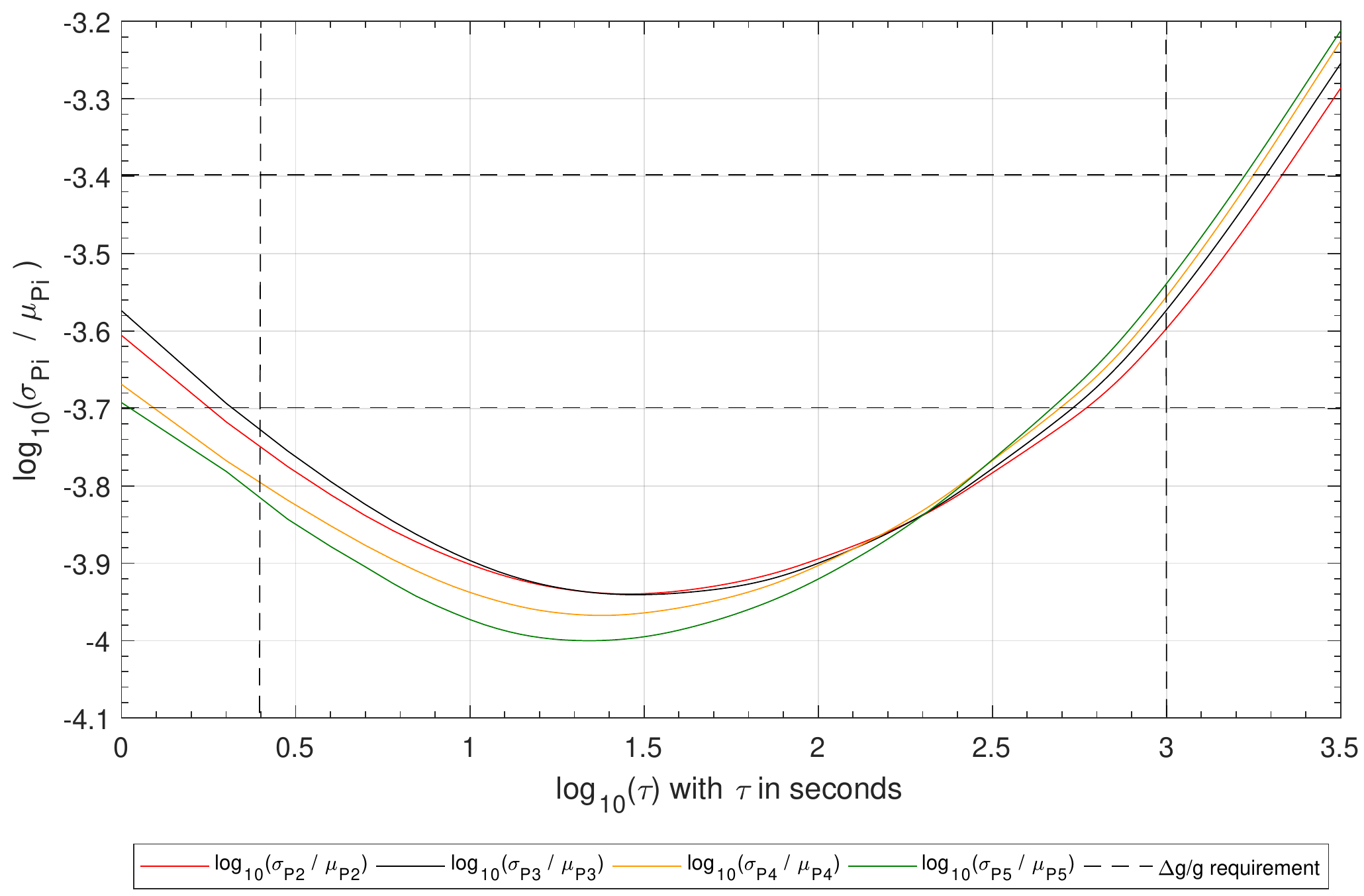}
\caption {Single channel Allan Standard Deviations. The two horizontal dashed lines represent single channel gain stability requirement of 2 - 4 $\times$ 10$^-$$^4$. The two vertical lines represent the timescale $\tau$ = 2.5 to 10$^3$ sec over which the gain stability is required.}
\label{SinglASD}
\end{figure}

The ASD for Channels 2 to 5 was calculated using equations \ref{ASDSC} and \ref{MSC}, and the results are shown in Figure \ref{SinglASD}. The two horizontal dashed lines represent the single channel gain stability requirement of 2 - 4 $\times$ 10$^-$$^4$. The two vertical lines represent the timescale $\tau$ = 2.5 - 10$^3$ sec over which the gain stability is required. Figure \ref{SinglASD} shows that the gain stability for Channels 2 to 5 is $<$ 2 $\times$ 10$^-$$^4$ over $\tau$ = 2.5 -$10^{2.65}$ sec and within 2 - 4 $\times$ 10$^-$$^4$ from $\tau$ = $10^{2.65}$ - $10^{3}$ sec, so an improvement from $\tau$ = $10^{2.65}$ - $10^{3}$ sec  is desirable. 

\subsubsection{Channel difference Allan Standard Deviations}
\begin{figure} [!htb]
\centering
\includegraphics[width=12cm, height = 6cm, keepaspectratio]{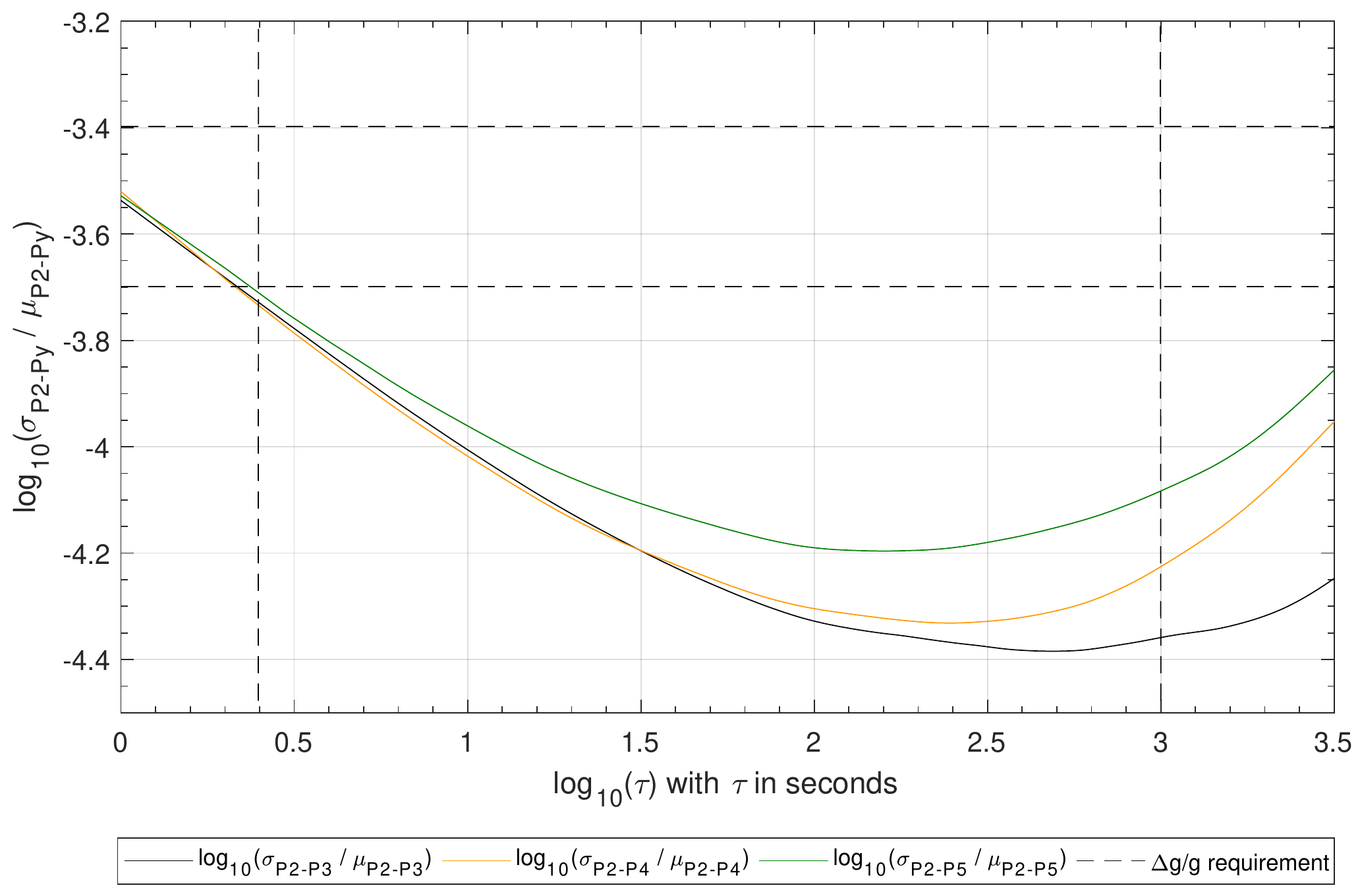}
\caption {Channel 2 Difference Allan Standard Deviations: Ch 2-3, Ch 2-4, and Ch 2-5.}
\label{Ch2diff}
\end{figure}

\begin{figure} [!htb]
\centering
\includegraphics[width=12cm, height = 6cm, keepaspectratio]{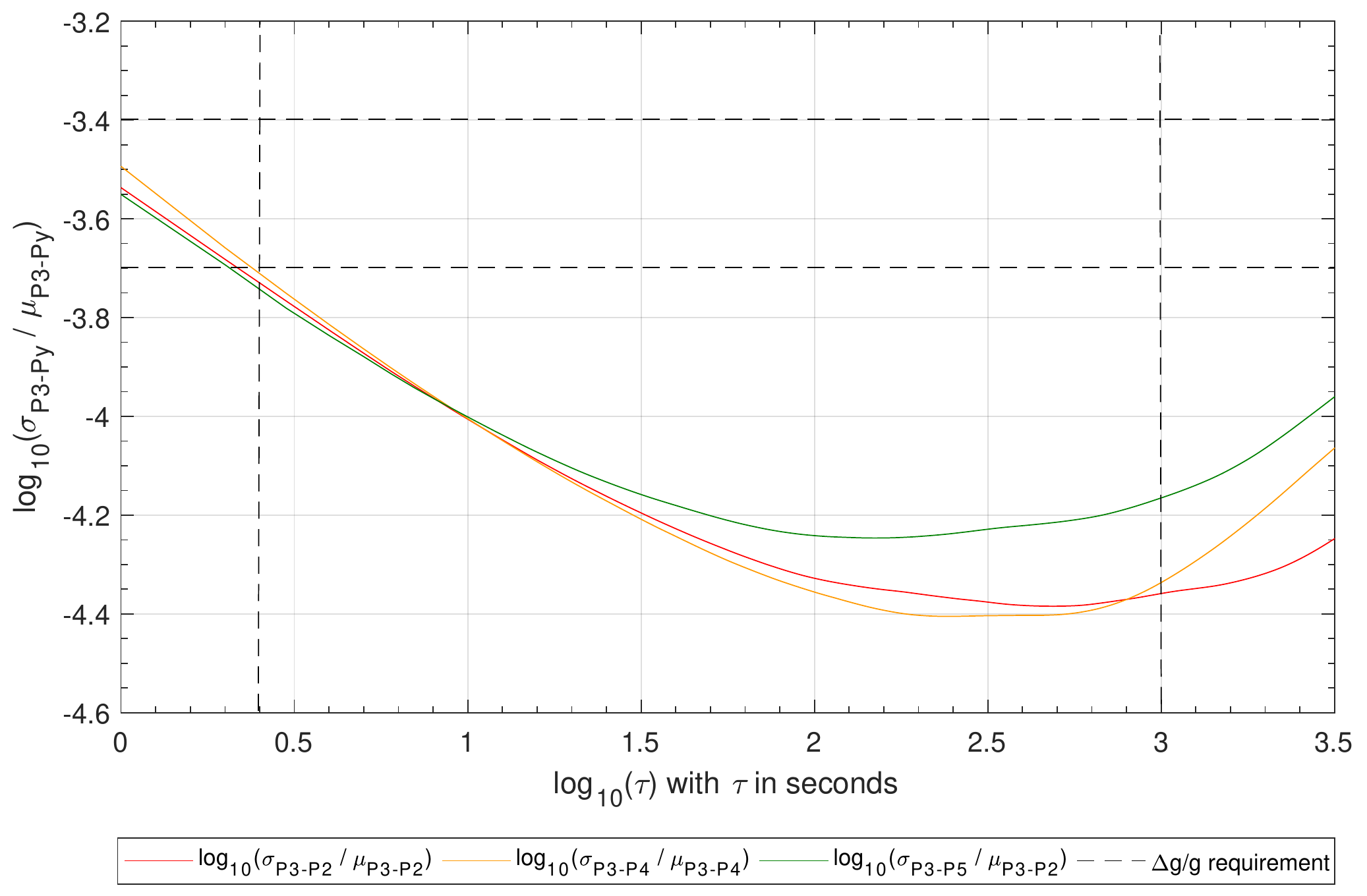}
\caption {Channel 3 Difference Allan Standard Deviations: Ch 3-2, Ch 3-4, and Ch 3-5.}
\label{Ch3diff}
\end{figure}

\begin{figure} [!htb]
\centering
\includegraphics[width=12cm, height = 6cm, keepaspectratio]{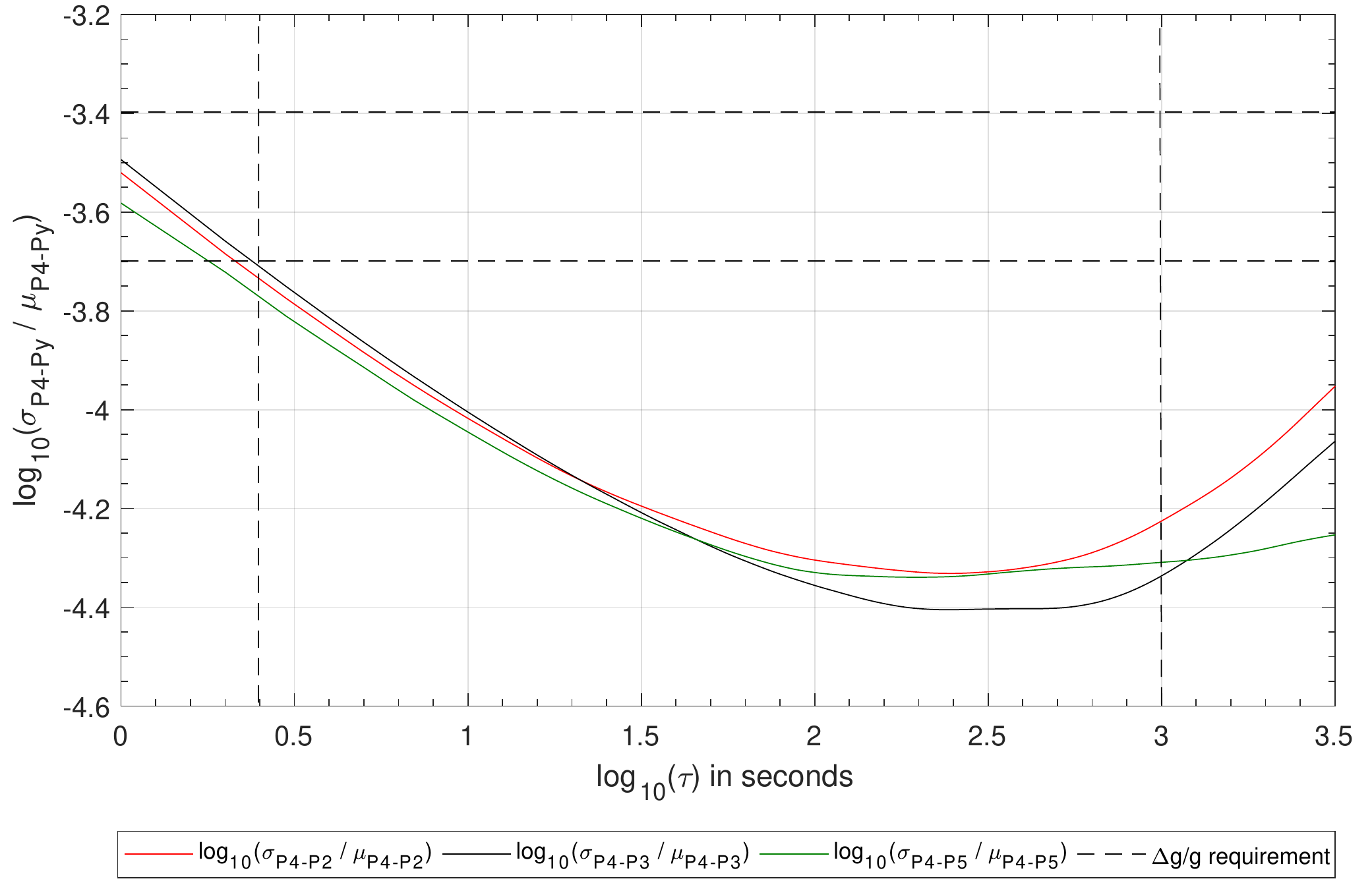}
\caption {Channel 4 Difference Allan Standard Deviations: Ch 4-2, Ch 4-3, and Ch 4-5.}
\label{Ch4diff}
\end{figure}

\begin{figure} [!htb]
\centering
\includegraphics[width=12cm, height = 6cm, keepaspectratio]{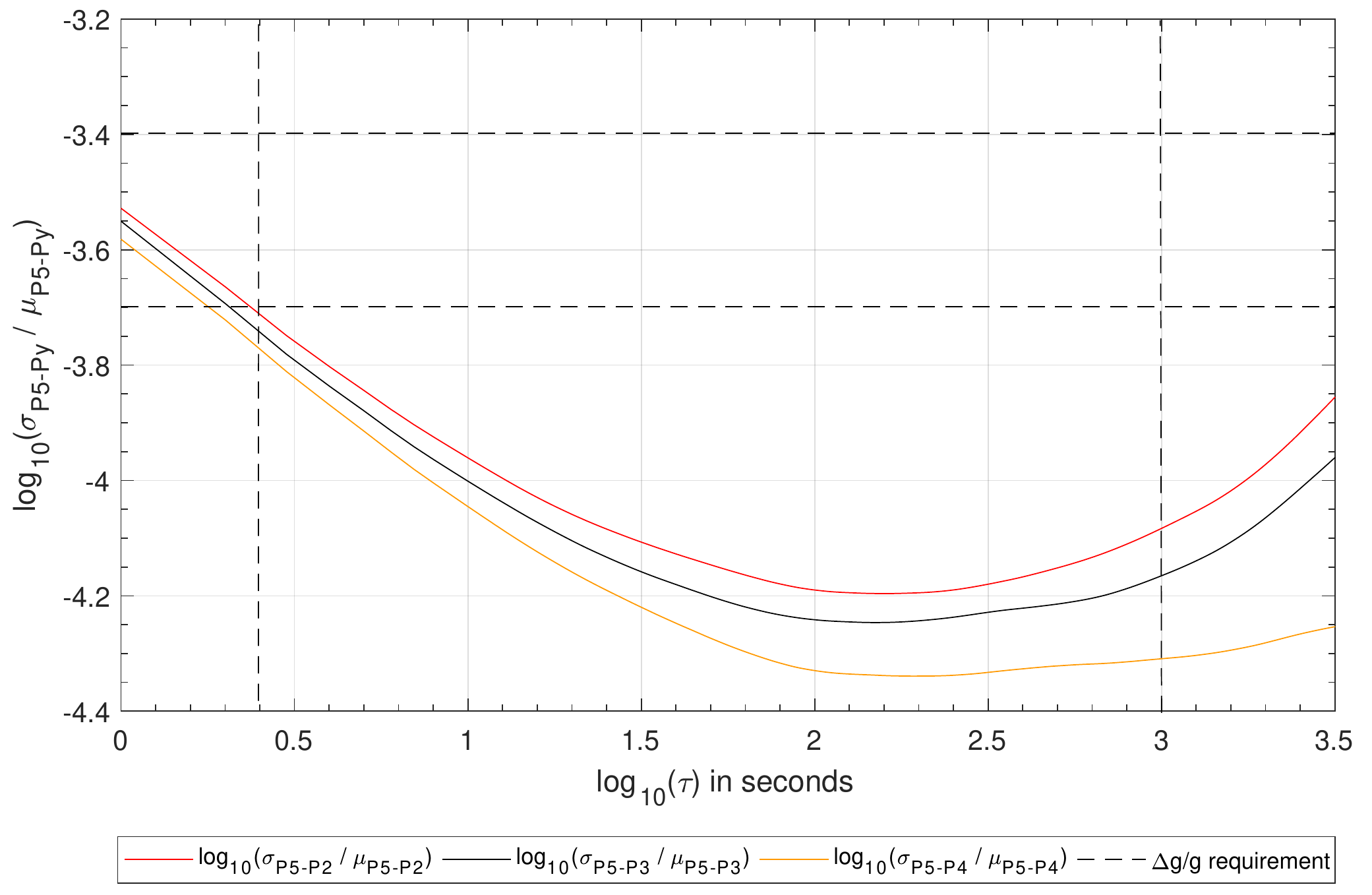}
\caption {Channel 5 Difference Allan Standard Deviations: Ch 5-2, Ch 5-3, and Ch 5-4.}
\label{Ch5diff}
\end{figure}

The channel difference ASDs are also calculated, as the single channel ASDs are limited by the stability of the input noise diode source and common signal path elements, whereas the difference ASDs measure the intrinsic stability of the independent signal paths. The channel difference ASDs were calculated for Channels 2 to 5 with respect to each other, and the results are shown in Figures \ref{Ch2diff}, \ref{Ch3diff}, \ref{Ch4diff}, and \ref{Ch5diff}. The results show that the gain stability of 2 - 4 $\times$ 10$^-$$^4$ over $\tau$ = 2.5 - 10$^3$ sec is successfully met for all cases.  

\subsubsection{Observable Allan Standard Deviation}
\begin{figure} [!htb]
\centering
\includegraphics[width=12cm, height = 6cm, keepaspectratio]{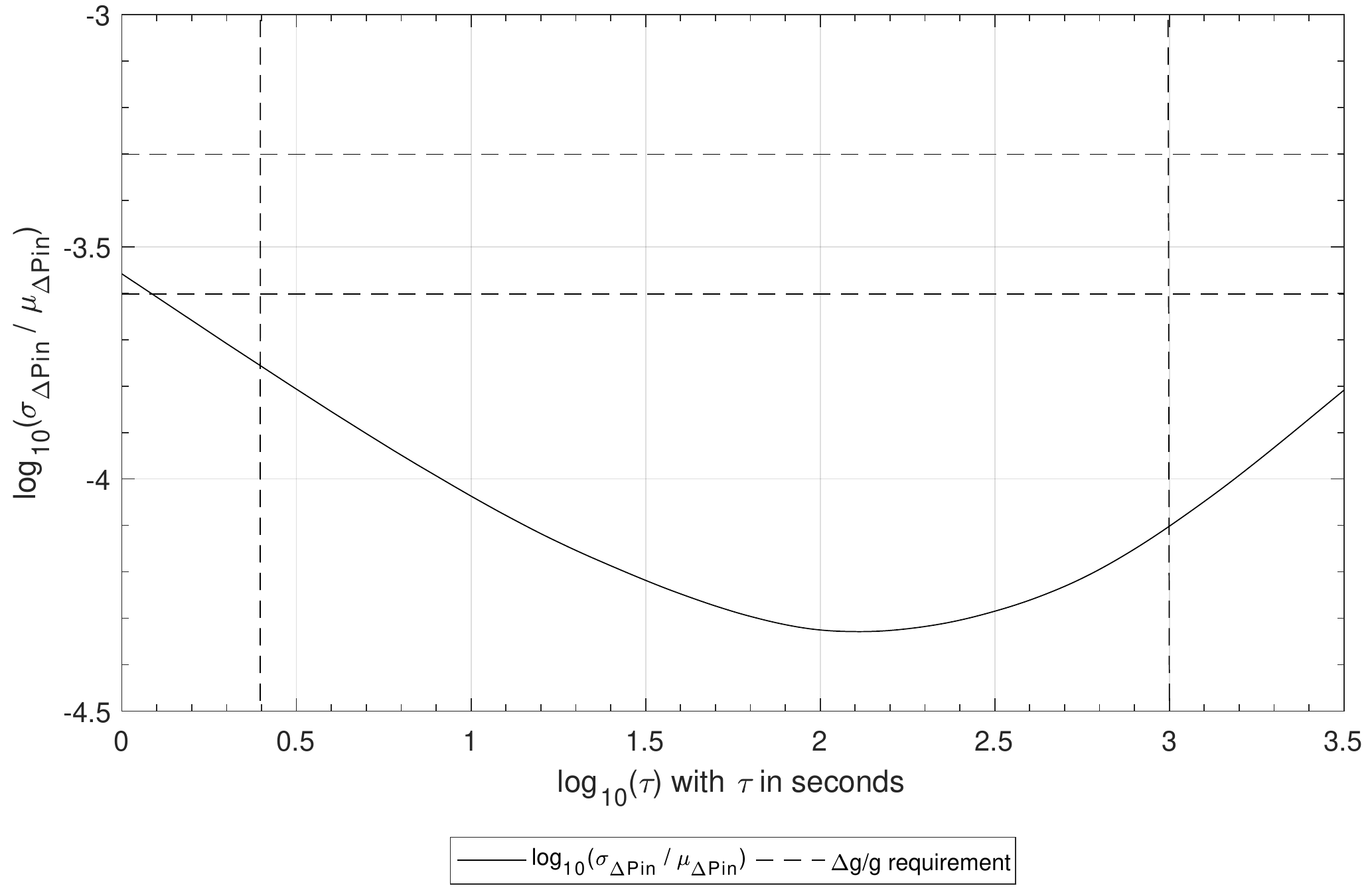}
\caption {Observable $\Delta$P$_{in}$ Allan Standard Deviation.}
\label{ObsASD}
\end{figure}

The ASD of \emph{$\Delta$P$_{in}$} was calculated using  equations \ref{ASDobs} and \ref{Mobs}, and the result is shown in Figure \ref{ObsASD}. The result shows that the gain stability requirement of 2.5 - 5 $\times$ 10$^-$$^4$ over $\tau$ = 2.5 - 10$^3$ sec is successfully met.

\subsubsection{Temperature Allan Standard Deviations}
\begin{figure} [!h]
\centering
\includegraphics[width=12cm, height = 6cm, keepaspectratio]{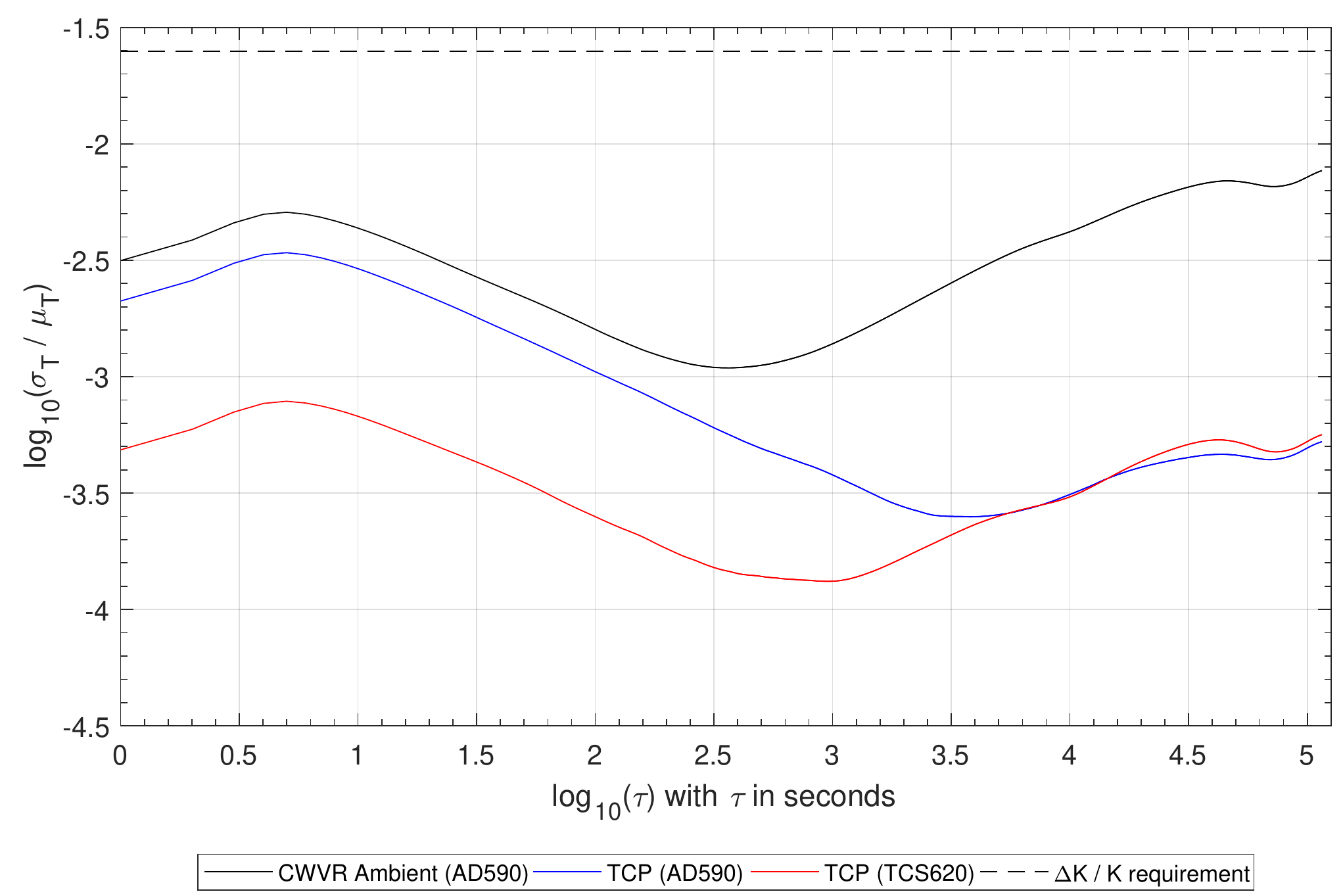}
\caption {Temperature Allan Standard Deviations. The horizontal dashed line is the 25 mK requirement.}
\label{TempASD}
\end{figure}

The ASD of the three temperature sensors was calculated, and the results are shown in Figure \ref{TempASD}. The results show that the temperature stability requirement of $\sim$ 25 mK was met for all the thermistors. The ASD of the TCP AD590 follows that of the TCP TCS-620 on longer timescales as expected. The measuring circuit employed by the TCS620 has lower noise for short time scales, and is more representative of the true temperature stability of the TCP.  

\subsection{Gain stability with temperature correction}
\subsubsection{Temperature correlation}
\begin{figure} [!htb]
\centering
\includegraphics[width=12cm, height = 6cm, keepaspectratio]{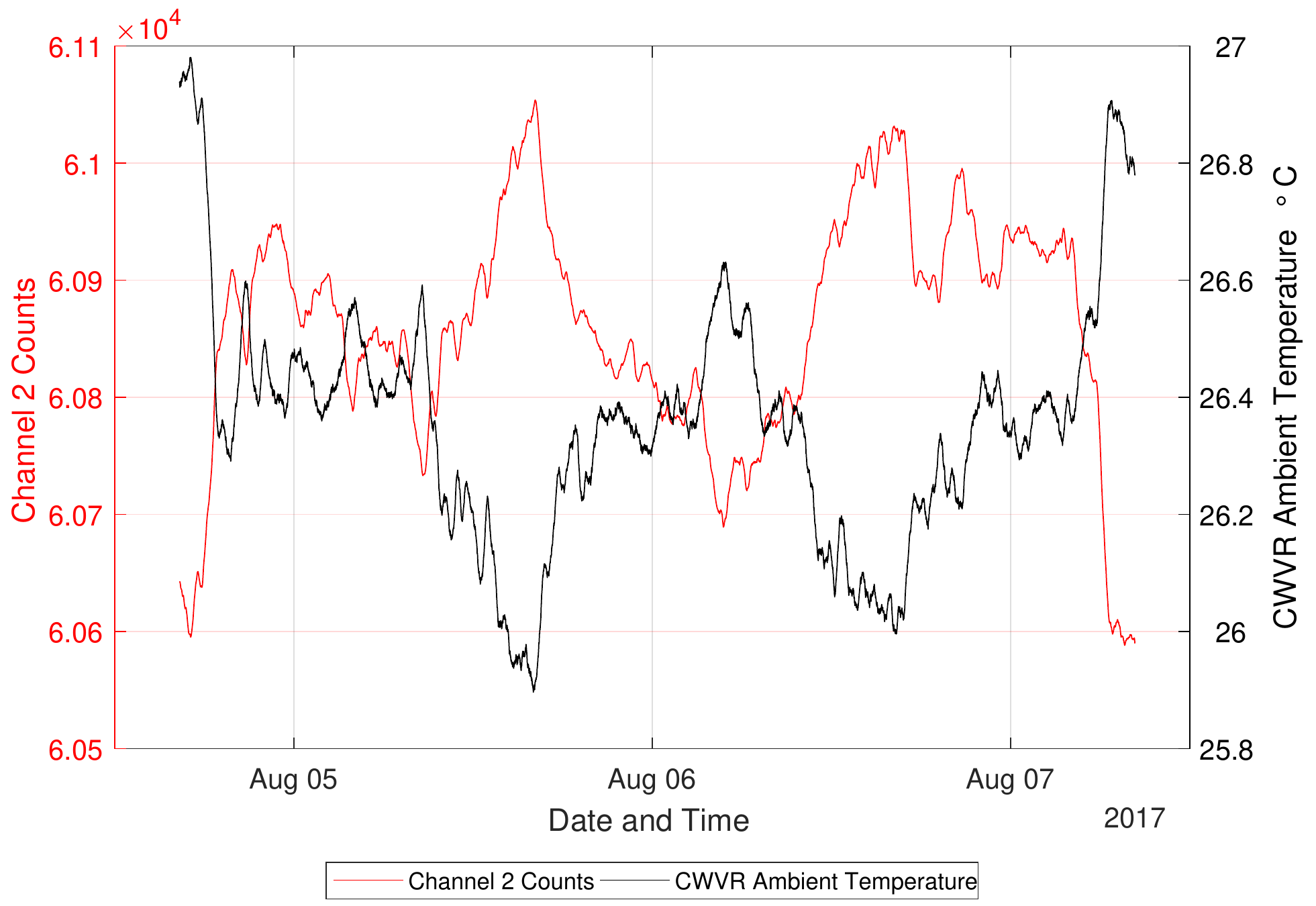}
\caption {Ch 2 Counts \& CWVR Ambient Temperature over 64 hr period averaged with a 20 min running mean.}
\label{Correlation}
\end{figure}

Figure \ref{Ch2-5} shows that the large-scale fluctuations in the output counts are consistent for Channels 2 to 5, suggesting a common source of the fluctuations. The CWVR ambient temperature and Channel 2 counts averaged with a 20 min running mean are plotted as a function of time over the 64 hr period in Figure \ref{Correlation}. It is apparent from Figure \ref{Correlation} that the CWVR ambient temperature and the output counts are negatively correlated. 

\begin{figure} [!htb]
\centering
\includegraphics[width=12cm, height = 7cm, keepaspectratio]{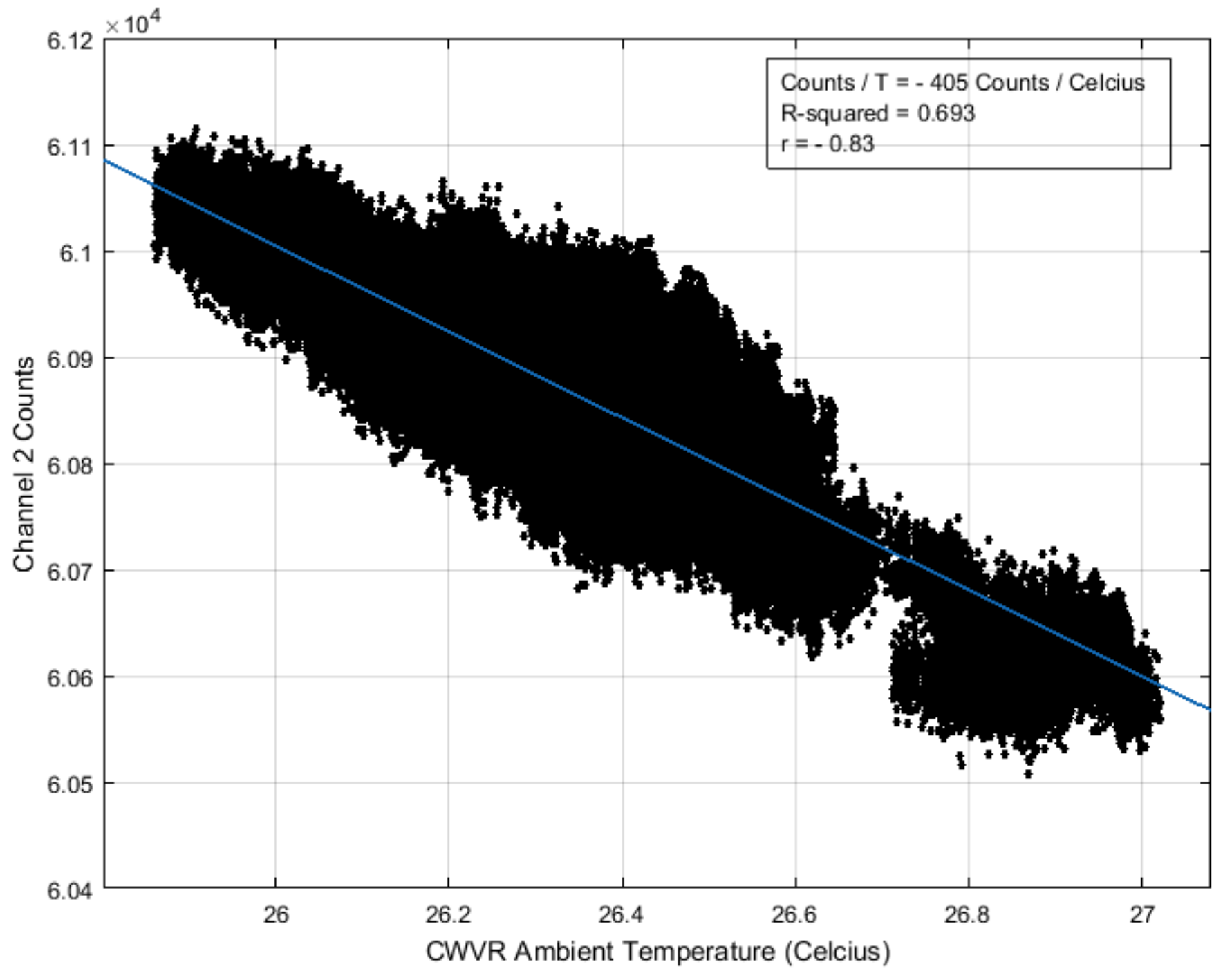}
\caption {Scatter plot of Channel 2 counts and CWVR ambient temperature.}
\label{Scatter}
\end{figure}

A scatter plot of Channel 2 counts versus CWVR ambient temperature smoothed with a 5 min running mean is shown in Figure \ref{Scatter}. The Pearson correlation coefficient between the Channel 2 counts and the CWVR ambient temperature is calculated to be r = -0.83 with 99.00$\%$ confidence. The coefficient of determination, R$^2$, for the linear fit is 0.69, providing an indication of the goodness of the fit. The slope of the fit indicates a change of $\sim$ 405 counts per 1\degree C change in the CWVR ambient temperature.

\subsubsection{Temperature correction}

\begin{figure} [!htb]
\centering
\includegraphics[width=12cm, height = 6cm, keepaspectratio]{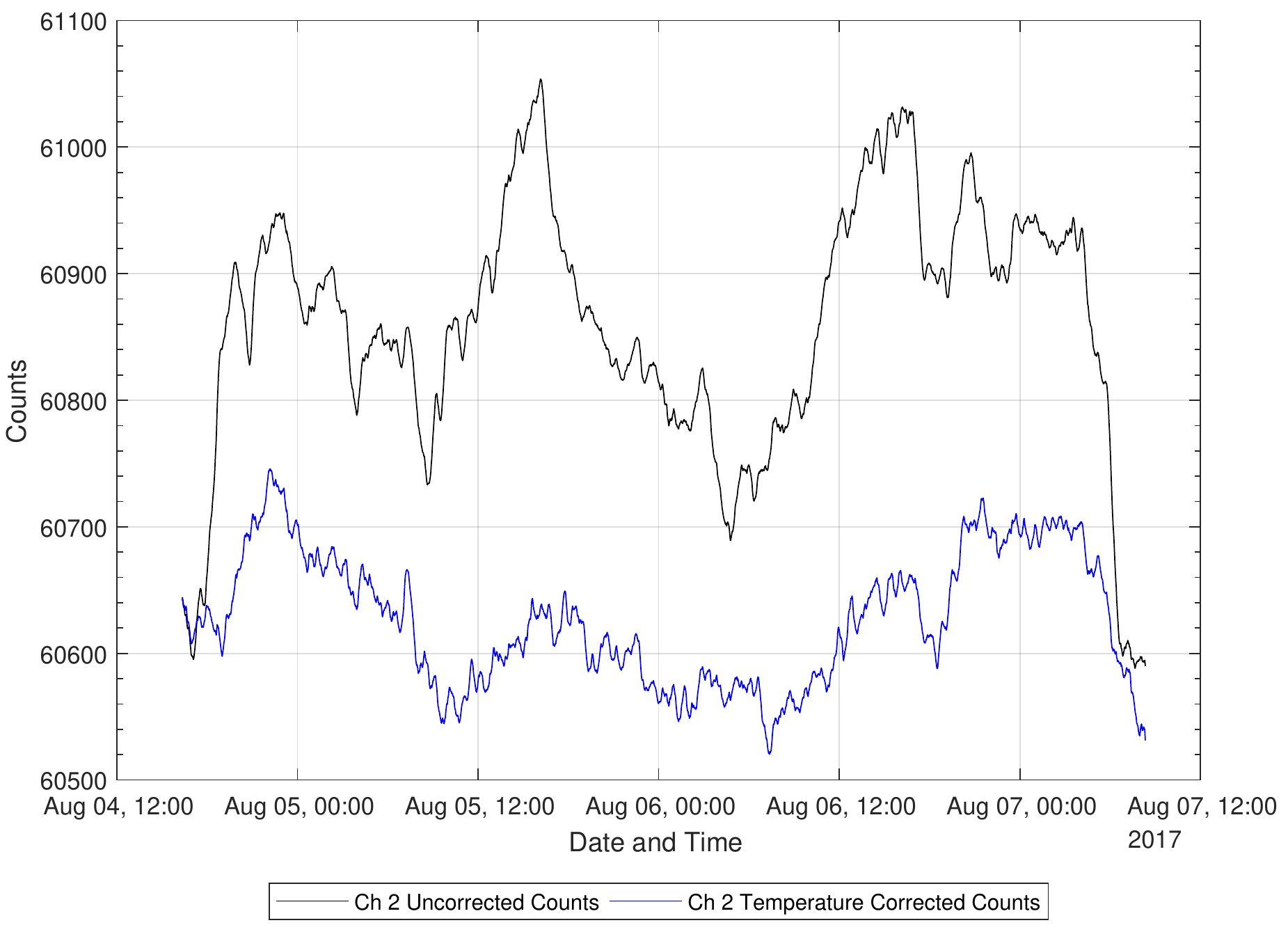}
\caption {Channel 2 uncorrected and temperature corrected counts with n = 5 min over 64 hr. The plots are averaged with a 20 min running mean.}
\label{CorrPlot}
\end{figure}

It is possible to correct for the changes in counts due to changes in the CWVR ambient temperature. The temperature data is typically  noisy over short timescales, so averaging over short timescales is necessary to effectively apply the correction. Equations \ref{Correct1}, \ref{Corr2}, and \ref{Corr3} are used to  apply the correction.

\begin{equation} \label{Correct1}
  Count_{corr,i} =
  \begin{cases}
    Count_{i} & \textit{for $i = 1 , 2 , 3 ,..., \;$n}\\
    Count_{i} - A(T_{sm,i} - T_{ave}) & \textit{for $i = n+1,n+2,..., \;$N} \\
     \end{cases}
\end{equation}

\begin{equation} \label{Corr2}
  T_{ave} = \frac {1} {n} \sum_{i=1}^{n} T_{i}
\end{equation}
\begin{equation} \label{Corr3}
  T_{sm,i} = \frac {1} {n} \sum_{j=i-n}^{i} T_{j}
\end{equation}

\noindent where \emph{A} = -405, \emph{Count$_{corr,i}$} is the temperature corrected count at point \emph{i}, \emph{Count$_{i}$} is the measured count at point \emph{i}, \emph{T$_{ave}$} is the mean of the CWVR ambient temperature for the first \emph{n} seconds of a \emph{N}-second total observation time, \emph{T$_{sm,i}$} is the temperature at point \emph{i} taken as the mean from the preceding \emph{n} seconds to point \emph{i}, and \emph{T$_{j}$} is the measured CWVR ambient temperature at point \emph{j}. The result of applying temperature correction with \emph{n} = 5 min for Channels 2 is shown in Figure \ref{CorrPlot}, and there is a clear improvement in the large-scale count fluctuations. Channels 3 to 5 show a similar improvement.

\subsubsection{Single channel Allan Standard Deviations with temperature correction}

\begin{figure} [!htb]
\centering
\includegraphics[width=12cm, height = 6cm, keepaspectratio]{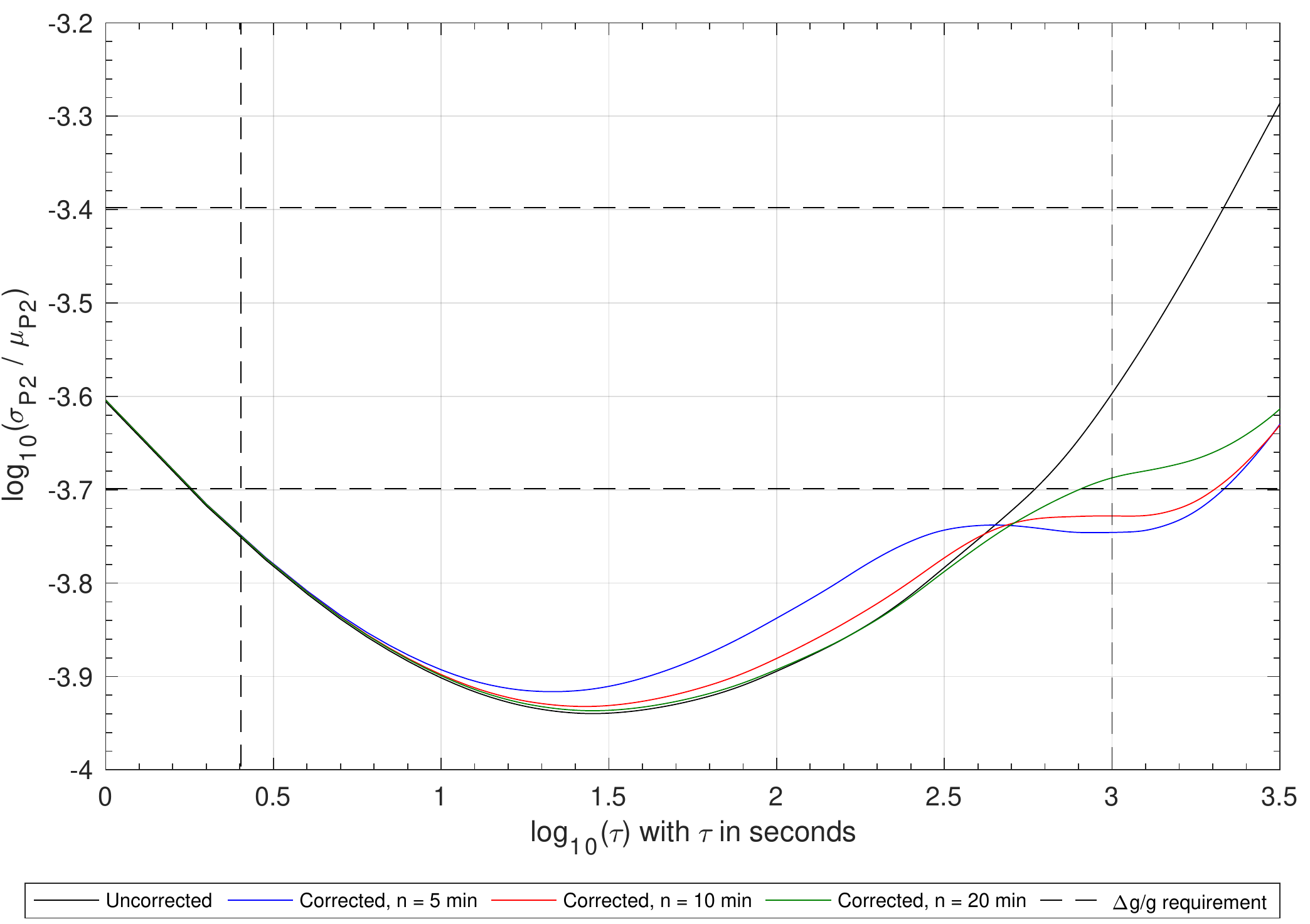}
\caption {Channel 2 Temperature Corrected Allan Standard Deviations for n = 5, 10, and 20 min.}
\label{Ch2ASDCorr}
\end{figure}

The temperature corrected and uncorrected ASDs for Channels 2 for n = 5, 10, and 20 min are shown in Figure \ref{Ch2ASDCorr}. The results show that while the temperature corrected ASDs are more stable than the uncorrected ASDs at $\tau$$_{long}$ = 10$^3$ sec, the uncorrected ASDs provide better stability from $\tau$ $\sim$ 10$^{0.8}$ - 10$^{2.6}$ sec. Temperature averaging time, n = 5 min, results in slightly worse gain stability from $\tau$ $\sim$ 10$^{0.8}$ - 10$^{2.65}$ sec but better gain stability at $\tau$$_{long}$ = 10$^3$ sec than the n = 10 and 20 min cases. The results indicate that a temperature averaging time of n = 10 min is optimal for stability at both short and long timescales, and the temperature corrected ASDs with n = 10 min are $<$ 2 $\times$ 10$^-$$^4$ over $\tau$ = 2.5 - 10$^3$ sec, which meet the requirement. Channels 3 to 5 show a similar improvement.  

\subsubsection{Observable Allan Standard Deviation with temperature correction}

\begin{figure} [!htb]
\centering
\includegraphics[width=12cm, height = 6cm, keepaspectratio]{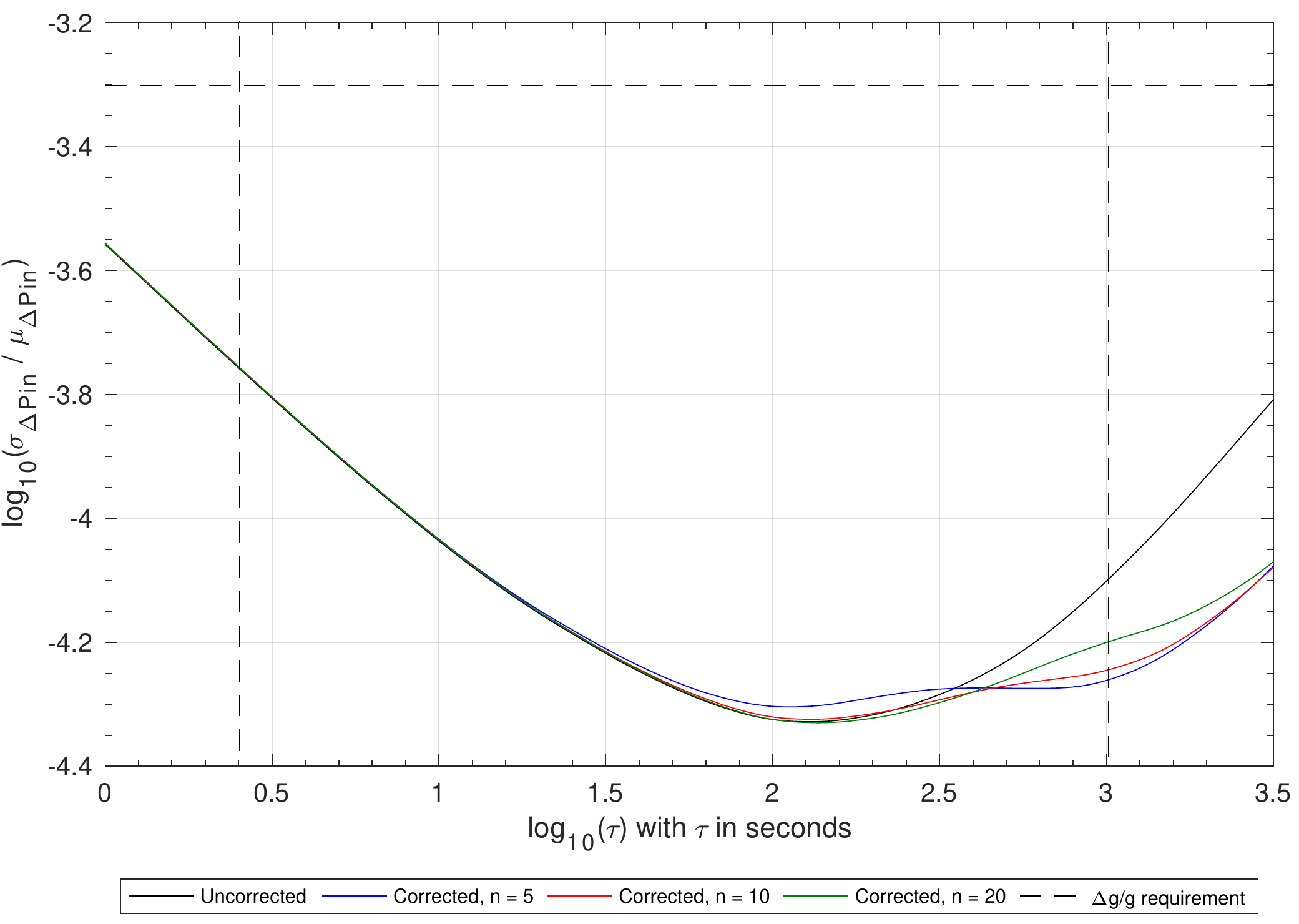}
\caption {Observable $\Delta$\textit{P}$_{in}$ temperature corrected Allan Standard Deviation for n = 5, 10, and 20 min.}
\label{DPinASDC}
\end{figure}

The temperature corrected and uncorrected ASDs of the observable for n = 5, 10, and 20 min are shown in Figure \ref{DPinASDC}. The results show that for a temperature averaging time of n = 10 min, the temperature corrected ASD is unaffected from $\tau$ = 10$^{0.4}$ - 10$^{2.4}$ sec and improves from $\tau$ = 10$^{2.4}$ - 10$^{3}$  sec  compared to the uncorrected ASD. Therefore, it is recommended that n = 10 min be used as the optimum temperature averaging time for temperature correction of the output counts.

\section{Summary}
The compact water vapor radiometer (CWVR) was characterized in the laboratory, and results show that the design meets the dynamic range, channel isolation, and gain stability requirements to be tested on an antenna. The channel isolation requirement of $<$ -20 dB was met, indicating $<$ 1\% power leakage between any two channels. Channel 1 needs to be repaired. The fluctuations in output counts are negatively correlated to the CWVR ambient temperature, with a change of $\sim$ 405 counts per 1\degree C change in temperature. With temperature correction, the single channel and channel difference gain stability is $<$ 2 $\times$ 10$^-$$^4$, and the observable gain stability is $<$ 2.5 $\times$ 10$^-$$^4$ over $\tau$ = 2.5 - 10$^3$ sec, all of which meet the gain stability requirement. Future work consists of building more CWVRs and testing the phase correlations on the VLA antennas to evaluate the use of WVR for not only the VLA, but also the Next Generation Very Large Array (ngVLA).  

\section*{Acknowledgments}
This work was possible through the National Radio Astronomy Observatory Graduate Student Research Assistantship program. I would like to thank R. Selina, B. Butler, R. Perley, J. Jackson, W. Grammer, and B. Willoughby for their guidance and support, C. Hennies for his assistance with the hardware, W. Koski for his assistance with the F318 module, the monitor and control system, and the 10 Hz calibration signal, G. Peck for the VHDL code, H. Frej for his assistance with the software, D. Urbain for his assistance with the K-band Dewar setup for the stability tests, and M. Morgan for the MMIC design. The National Radio Astronomy Observatory is a facility of the National Science Foundation operated under cooperative agreement by Associated Universities, Inc.


\begin{thebibliography}{}

\bibitem[Butler(1999)] {B1999} Butler, B. [1999] {\it Some Issues for Water Vapor Radiometry at the VLA}, VLA Scientific Memo $\#$ 177.
\bibitem[Carilli \& Holdaway(1999)] {Carilli} Carilli, C.~L \& Holdaway, M.~A. [1999] {\it Tropospheric phase calibration in millimeter interferometry}, {\it Radio Science}, vol. 34, no. 4, pp. 817-810. 
\bibitem[Chandler {\it et al.}(2004a)] {Chandlera} Chandler, C.~J, Brisken, W.~F, Butler, B.~J, Hayward, R.~H, M., Willoughby, B.~E. [2004] {\it Results of Water Vapor Radiometry Tests at the VLA}, EVLA Memo $\#$ 73.
\bibitem[Chandler {\it et al.}(2004b)] {Chandlerb} Chandler, C.~J, Brisken, W.~F, Butler, B.~J, Hayward, R.~H, Morgan, M., Willoughby, B.~E. [2004] {\it A Proposal to Design and Implement a Compact Water Vapor Radiometer for the EVLA}, EVLA Memo $\#$ 74.
\bibitem[Clark(2015)] {Clark2015} Clark, B. [2015] {\it Calibration Strategies for the Next Generation VLA}, ngVLA Memo $\#$ 2.
\bibitem[Cornwell \& Fomalont(1999)] {CF1999} Cornwell, T. J., and Fomalont, E. [1999] {\it Aperture Synthesis in Radio Astronomy $\rom{2}$}, edited by Taylor, G., Carilli, C., \& Perley, R., pp. 187-199, stron. Soc. of the Pac., San Francisco, Calif.
\bibitem[Desai(1993)] {Desai1993} Desai, K. [1993] {\it Measurement of turbulence in the interstellar medium}, Ph.D. thesis, 89 pp., Univ. of Calif. at Santa Barbara.
\bibitem[Dicke {\it et al.}(1946)] {Dicke1946} Dicke, R. H., Beringer, R., Kyhl, R. L., and Vane, A. B. [1946] {\it Atmospheric absorption measurements with a microwave radiometer}, \emph{Phys. Rev., 70}, pp. 340-348.
\bibitem[Fomalont \& Perley(1999)] {FP1999} Fomalont, E., and Perley, R. A. [1999] {\it Aperture Synthesis in Radio Astronomy \rom{2}}, edited by Taylor, G., Carilli, C., and Perley, R. A. pp. 79-109, Astron. Soc. of the Pac., San Francisco, Calif.  
\bibitem[Holdaway(1992)] {Hold1992} Holdaway, M. A. [1992] {\it Possible phase calibration schemes for the MMA}, Millimeter Array Memo. 84, pp. 14.
\bibitem[Holdaway \& Owen(1995)] {HO1995} Holdaway, M. A., and Owen F. N. [1995] {\it A test of fast switching phase calibration with the VLA at 22GHz}, Millimeter Array Memo. 126, pp. 8.
\bibitem[Koski(2017)] {Koski2017} Koski, W. M. [2017] {\it F318 WVR \& Low-Band Interface Module}, EVLA Front End Group, Document $\#$ A23185N0044, Revision B.
\bibitem[Lay(1997)] {Lay1997}  Lay, O. P. [1997] {\it Phase calibration and water vapor radiometry for millimeter-wave arrays}, \emph{Astron. Astrphys. Suppl. Ser., 122}, pp. 547-565.
\bibitem[Sterling {\it et al.}(2004)] {Sterling2004}  Sterling, A., Hills, R., Richer, J., Pardo, J. [2004] {\it 183 GHz water vapour radiometers for ALMA: Estimation of phase erros under varying atmospheric conditions}, ALMA Memo $\#$ 496.
\bibitem[Sutton \& Hueckstaedt(1997)] {SH1997}   Sutton, E. C., and Hueckstaedt, R. M. [1997] {\it Radiometric monitoring of atmospheric water vapor as it pertains to phase correction in millimeter interferometry}, \emph{Astron. Astrophys. Suppl. Ser., 119}, pp. 559-567.
\bibitem[Thompson {\it et al.}(2001)] {TS2001} Thompson, A.~R, Moran, J.~M, Swenson Jr., G.~W [2001] {\it Interferometry and Synthesis in Radio Astronomy}, 2nd edition (John Wiley \& Sons, New York). 


\end{thebibliography}
\end{document}